\documentclass[runningheads]{llncs}
\usepackage{xcolor}

\usepackage{graphicx}
\usepackage{amsmath}
\usepackage{amssymb}
\usepackage{booktabs}
\usepackage{adjustbox}
\begin{document}

\title{Improved Super Resolution of MR Images Using CNNs and Vision Transformers}

\author{Dwarikanath Mahapatra }

%\authorrunning{*****************}
\institute{Inception Institute of AI, UAE }

\maketitle

% \midlauthor{\Name{Dwarikanath Mahapatra\nametag{$^{1,2}$}} \Email{dwarikanath.mahapatra@inceptioniai.org}\\
% \addr $^{1}$ Inception Institute of AI, Abu Dhabi, UAE \\
% \addr $^{2}$ Faculty of Engineering, Monash University, Melbourne, Australia
% \AND
% \Name{Zongyuan Ge\nametag{$^{2,3,4}$}} \Email{zongyuan.ge@monash.edu}\\
% \addr $^{3}$ Airdoc-Monash Research Australia \\
% \addr $^{4}$ Monash eResearch Centre Australia \\
% }

\begin{abstract}
   State of the art magnetic resonance (MR) image super-resolution methods (ISR) using convolutional neural networks (CNNs) leverage limited contextual information due to the limited spatial coverage of CNNs. Vision transformers (ViT) learn better global context that is helpful in generating superior quality HR images. We combine local information of CNNs and global information from ViTs for image super resolution and output super resolved images that have superior quality than those produced by state of the art methods. We include  extra constraints through multiple novel loss functions that preserve structure and texture information from the low resolution to high resolution images.


\end{abstract}

% \begin{keywords}
% MRI, super resolution, disentanglement, CNN, ViT
% \end{keywords}

\section{Introduction}

Image super-resolution (ISR) takes low resolution (LR) image inputs and reconstructs its corresponding high resolution (HR) version thus enabling detailed examination of  interesting regions. 
This is particularly relevant for medical image analysis where physics of the imaging systems limits the spatial resolution of radiological images (e.g. MRI, Xray) since obtaining HR images requires longer scanning time, and leads to lower signal-to-noise ratio and smaller spatial coverage \cite{ZhangCVPR21_32,Mahapatra_Media_SIBNET,MahapatraGZSLTMI,LieTMI_2022,Devika_IEEE,MonusacTMI,Mahapatra_Thesis}, \cite{KuanarVC,MahapatraTMI2021,JuJbhi2020,Frontiers2020,Mahapatra_PR2020,ZGe_MTA2019}, \cite{Behzad_PR2020,Mahapatra_CVIU2019,Mahapatra_CMIG2019,Mahapatra_LME_PR2017}, \cite{Zilly_CMIG_2016,Mahapatra_SSLAL_CD_CMPB,Mahapatra_SSLAL_Pro_JMI,Mahapatra_LME_CVIU}. HR images provide more detailed information about local structures and textures resulting in higher accuracy in disease diagnosis and planning \cite{ZhangCVPR21_5,LiTMI_2015,MahapatraJDI_Cardiac_FSL,Mahapatra_JSTSP2014}, \cite{MahapatraTIP_RF2014,MahapatraTBME_Pro2014,MahapatraTMI_CD2013}, \cite{MahapatraJDICD2013,MahapatraJDIMutCont2013,MahapatraJDIGCSP2013,MahapatraJDIJSGR2013,MahapatraTrack_Book}, \cite{MahapatraJDISkull2012,MahapatraTIP2012,MahapatraTBME2011},  \cite{MahapatraEURASIP2010,MahapatraTh2012,MahapatraRegBook}.
Since originally acquired LR images pose challenges for accurate analysis it is important to have a reliable ISR method.

Recent works  demonstrate the potential of 
 convolutional neural networks (CNNs) in generating HR images by using SRCNN  \cite{ZhangCVPR21_7,ZhangCVPR21_8,Souryaisbi22,mahapatra2022_midl}, \cite{Mahapatra_CVAMD2021,PandeyiMIMIC2021},  \cite{SrivastavaFAIR2021,Mahapatra_DART21b,Mahapatra_DART21a,LieMiccai21}, \cite{TongDART20,Mahapatra_MICCAI20,Behzad_MICCAI20,Mahapatra_CVPR2020}, \cite{Kuanar_ICIP19,Bozorgtabar_ICCV19,Xing_MICCAI19,Mahapatra_ISBI19,MahapatraAL_MICCAI18,Mahapatra_MLMI18}, \cite{Sedai_OMIA18,Sedai_MLMI18,MahapatraGAN_ISBI18,Sedai_MICCAI17,Mahapatra_MICCAI17}, residual learning in VDSR (Very Deep Super Resolution) \cite{VDSR,Roy_ISBI17,Roy_DICTA16,Tennakoon_OMIA16,Sedai_OMIA16,Mahapatra_OMIA16}, \cite{Mahapatra_MLMI16,Sedai_EMBC16,Mahapatra_EMBC16,Mahapatra_MLMI15_Optic}, \cite{Mahapatra_MLMI15_Prostate,Mahapatra_OMIA15,MahapatraISBI15_Optic,MahapatraISBI15_JSGR,MahapatraISBI15_CD}, \cite{KuangAMM14,Mahapatra_ABD2014,Schuffler_ABD2014,Schuffler_ABD2014_2,MahapatraISBI_CD2014}, \cite{MahapatraMICCAI_CD2013,Schuffler_ABD2013,MahapatraProISBI13,MahapatraRVISBI13}, and the information distillation network (IDN) \cite{ZhangCVPR21_15}. \cite{ZhangCVPR21_48,MahapatraWssISBI13,MahapatraCDFssISBI13},  \cite{MahapatraCDSPIE13,MahapatraABD12,MahapatraMLMI12,MahapatraSTACOM12}, \cite{VosEMBC,MahapatraGRSPIE12,MahapatraMiccaiIAHBD11,MahapatraMiccai11}, \cite{MahapatraMiccai10,MahapatraICIP10,MahapatraICDIP10a},  \cite{MahapatraICDIP10b,MahapatraMiccai08,MahapatraISBI08}, \cite{MahapatraICME08,MahapatraICBME08_Retrieve,MahapatraICBME08_Sal,MahapatraSPIE08,MahapatraICIT06}  leverage hierarchical features  in residual deep networks (RDN) while \cite{ZhangCVPR21_5} combine 3D dense networks and adversarial learning for MRI super resolution.
MR images have inherent  characteristics such as repeating structural patterns making them less complex than natural images. Secondly, they have a large proportion of background pixels. Since most approaches give the background and foreground equal importance it does not lead to good feature learning.
Also, CNN  methods capture mostly local context information  and do not explore the global aspects. Zhang et al. in \cite{ZhangCVPR21,UDA_Ar,IccvGZSl_Ar,ISR_MIDL_Ar,GCN_MIDL_Ar,DevikaAccess_Ar}, \cite{SouryaISBI_Ar,Covi19_Ar,DARTGZSL_Ar}, \cite{DARTSyn_Ar,Kuanar_AR2,TMI2021_Ar,Kuanar_AR1},  \cite{Lie_AR2,Lie_AR,Salad_AR}, \cite{Stain_AR,DART2020_Ar,CVPR2020_Ar},  \cite{sZoom_Ar,CVIU_Ar,AMD_OCT,GANReg2_Ar} propose a squeeze and excitation network to capture the global characteristics thus leading to improved super resolution output. However, squeeze and excitation relies on CNN features to capture global context which is not optimal.

Vision transformers (ViT) \cite{ViT} are an exciting new development that  effectively capture long range contextual information from images. Given sufficient training data ViTs have been shown to outperform state of the art CNN based methods for classification and segmentation. In this work we propose to combine CNNs and ViT for performing super resolution of MR images. CNNs learn local details while ViT captures the global context much better than previously proposed methods.
Inherent to the ViT is a self attention module that focuses on the 
important parts of the image and thus improves SR quality. Our method also uses feature disentanglement to improve super resolution. % 

\section{Related Work}

\textbf{MR Image Super-resolution:}
ISR has been widely applied to MR images \cite{ZhangCVPR21_36,ZhangCVPR21_27}, and spectroscopy MRI \cite{ZhangCVPR21_16,ZhangCVPR21_17,GANReg1_Ar,PGAN_Ar},  \cite{Haze_Ar,Xr_Ar,RegGan_Ar}, \cite{ISR_Ar,LME_Ar,Misc,Health_p,Pat2},  \cite{Pat3,Pat4,Pat5,Pat6,Pat7}.
Initial methods  achieved 
multiple frame image super resolution via alignment of multiple noisy LR images which proved to be very challenging \cite{ZhangCVPR21_49}. Recent deep learning based ISR approaches show
superior performance for MR image super resolution \cite{ZhangCVPR21_6,ZhangCVPR21_31,ZhangCVPR21_49} but use large models that pose challenges in real world settings. Zhang et al. in \cite{ZhangCVPR21} propose a squeeze and excitation attention network as part of a lightweight model for ISR. \cite{FengMulti} achieve multi contrast MRI super resolution using multi stage networks. \cite{HuMIDL21} use graph convolution networks for MRI super resolution, while in other related work recent methods have proposed hybrid-fusion networks for Multi-modal synthesis of MRI \cite{ZhouTMI20,Pat8,Pat9},  \cite{Pat10,Pat11,Pat12}, \cite{Pat13,Pat14,Pat15},  \cite{Pat16,Pat17,Pat18}, and  \cite{DarTMI} synthesize multi-contrast MRI using conditional GANs.

\textbf{Attention Mechanism:}
Attention mechanisms enables 
 adaptive resource allocation by focusing on important
image regions \cite{ZhangCVPR21_13} and are popular for many tasks like image recognition \cite{ZhangCVPR21_1} and image captioning \cite{ZhangCVPR21_44}, as well as ISR \cite{ZhangCVPR21_14,ZhangCVPR21_48}. They can be highly effective for MRI super resolution due to repeating patterns of relatively simpler structures and less informative background.

\textbf{Vision Transformers:}
Dosovitskiy et al. \cite{ViT}  demonstrate state-of-the-art performance on
image classification datasets using large-scale pre-training and
fine-tuning, and \cite{UNETR_5,UNETR_55} use ViTs for object detection. Hierarchical vision transformers with varying resolutions and spatial embeddings \cite{UNETR_30,UNETR_44} have been used to reduce feature resolution, while \cite{Esser} demonstrate success in high resolution image synthesis.%
 Recent work on transformer-based models for 
2D image segmentation include the SETR model that uses a pre-trained transformer
encoder with different CNN decoders \cite{Unetr52} for multi-organ segmentation in \cite{Unetr7}, and a transformer-based axial attention mechanism for 2D medical image segmentation \cite{Unetr41}.
Hatamizadeh et al. propose UNETR \cite{UNETR} for 3D medical image segmentation using transformers as the main encoder of a segmentation network and directly connecting to the decoder via skip connections.  
For 3D medical image segmentation,  \cite{Unetr47}
use a backbone CNN for
feature extraction, a transformer to process the encoded
representation and a CNN decoder for predicting
segmentation outputs, while \cite{Unetr43} 
use transformers in the bottleneck of a 3D encoder-decoder
CNN for semantic brain tumor segmentation. However, none of the methods use ViT for medical image super resolution.

\textbf{Motivation And Contribution:}
Context information is especially relevant for medical ISR since they provide additional cues to generate superior quality HR images. Our contributions are: 
1) We combine CNNs and ViTs for image super resolution. Local contextual cues from CNNs and global information from ViTs result in superior quality super resolved images than those produced by state of the art methods. A pre-trained ViT is finetuned using self supervised learning. 2) Using multiple loss functions we incorporate  extra constraints that preserve structural and semantic information in the generated super resolved image. 3) By comparing with results from \cite{HuMIDL21} we also demonstrate  our method's better ability to learn global features compared to graph based super resolution methods.

\section{Method}
\label{sec:met}

\subsection{Overview}

\begin{figure*}[t]
 \centering
\begin{tabular}{cc}
\includegraphics[height=4.6cm, width=7.5cm]{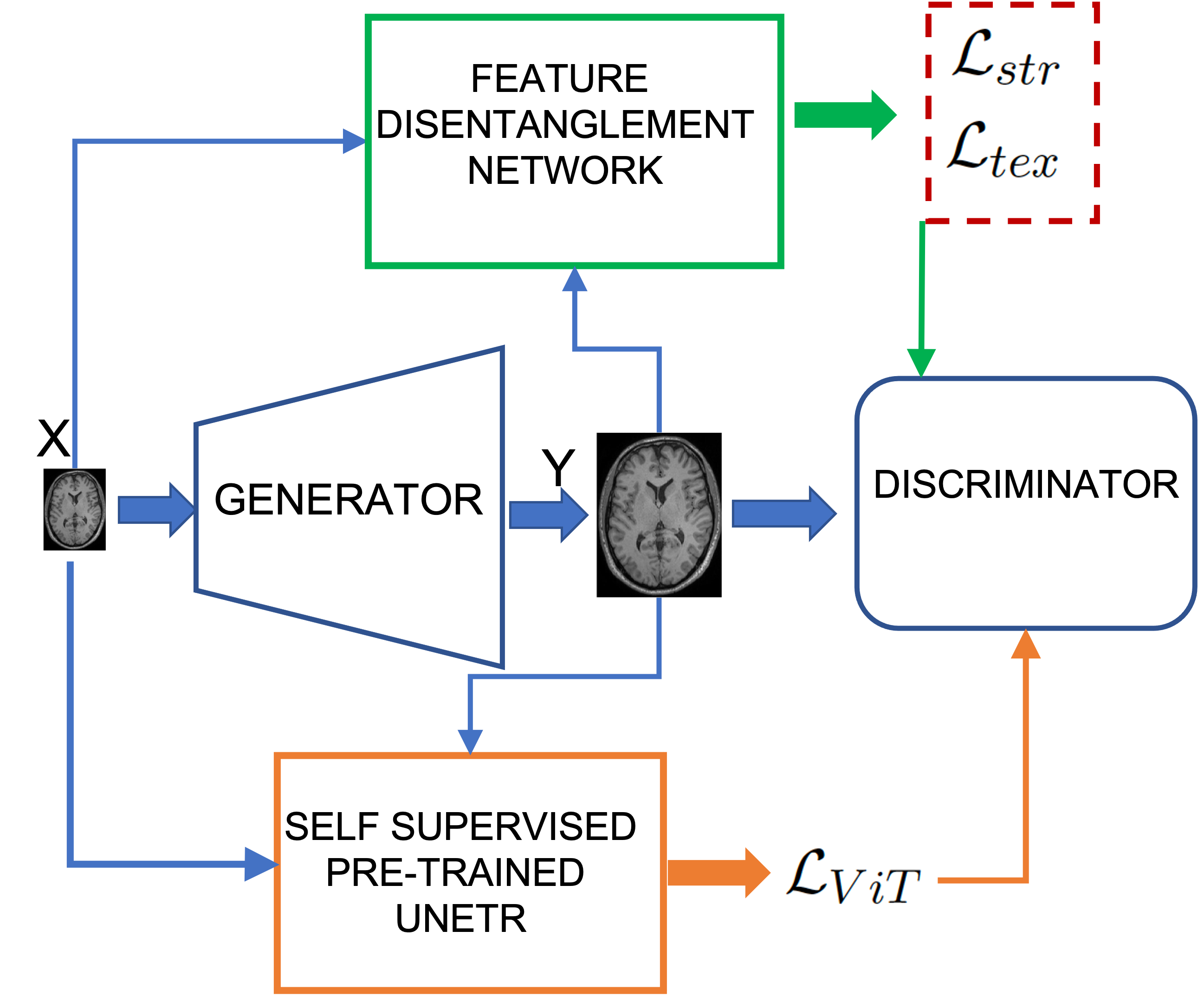} & 
\includegraphics[height=4.3cm, width=7.5cm]{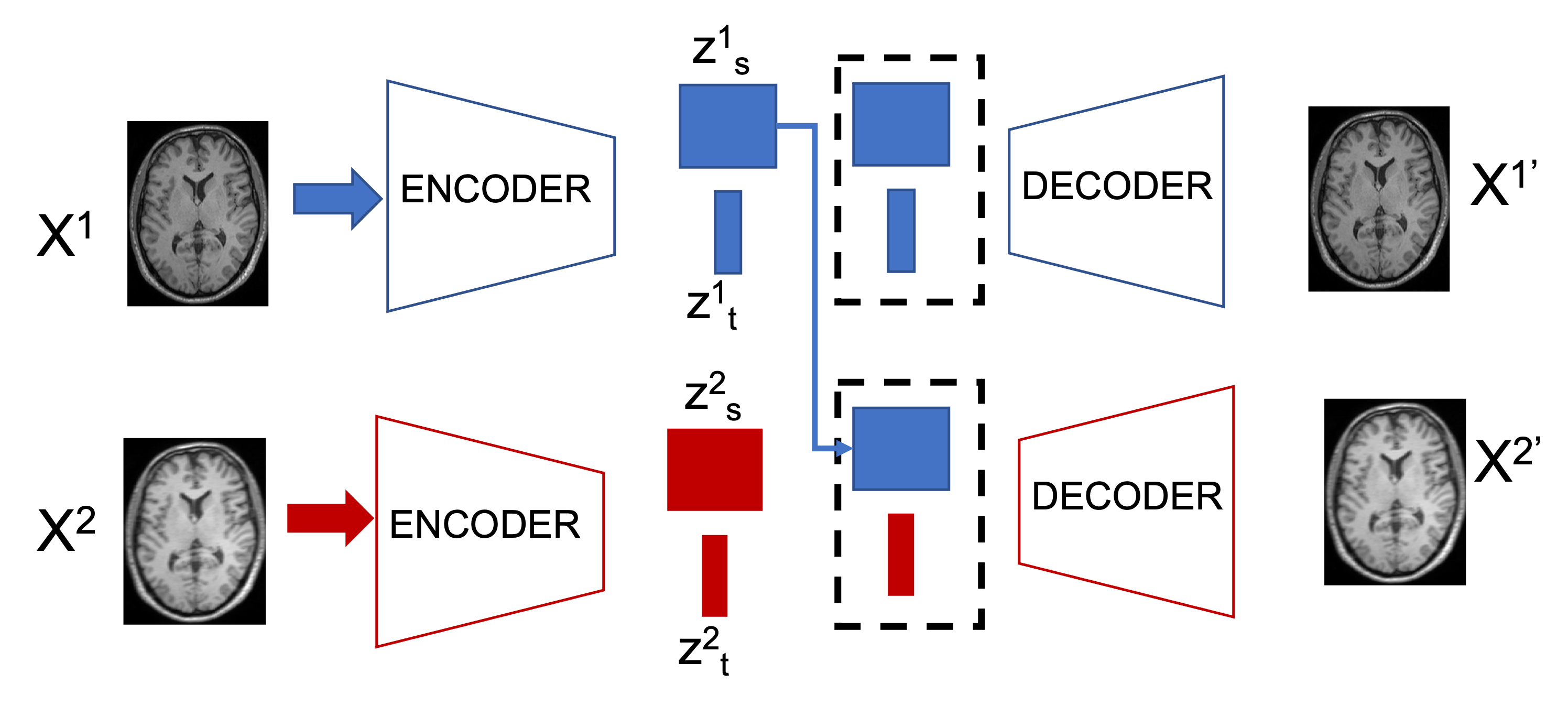} \\
(a) & (b)
\end{tabular}
\caption{(a) Workflow of our proposed method. LR images goes through a generator to get HR image, and multiple loss functions ensure that semantic information of the LR image is preserved in the HR image. (b) Architecture of feature disentanglement network using swapped autoencoders.}
\label{fig:workflow}
\end{figure*}

Given a low resolution (LR) image $x\in \mathcal{R}^{N\times N}$ our objective is to train a model that outputs a high resolution (HR) image $y\in \mathcal{R}^{M\times M}$, where $M>N$. 
Figure~\ref{fig:workflow} shows the workflow of our proposed method. 
The LR image $x$ goes through a generator network consisting of a series of convolution blocks and an upsampler that increases the image dimensions from $N\times N$ to $M\times M$. The discriminator module ensures $y$ satisfies the following constraints.
\begin{enumerate}
    \item The HR and LR image should have similar semantic characteristics since a higher resolution version should not alter image semantics. For this purpose we disentangle the image into structure and  and texture features, and ensure their respective semantic information is consistent across both images. %
    \item The HR image should preserve global and local context of the original LR image. To achieve it we use features extracted using a pre-trained ViT to effectively capture the relations in LR image and ensure this relationship is preserved in the HR image. 
\end{enumerate}

\subsection{Vision Transformers}

Vision transformers play an important role in our super resolution framework by serving as a robust and accurate feature extractor that integrates long range context and structural information.
We use the  ViT from UNETR  \cite{UNETR} pre-trained for MR image segmentation and fine tune it for our task. We briefly describe the architecture below (for full details please refer to Appendix~\ref{app:UNETR}) and also explain our modifications. %
UNETR uses the contracting-expanding pattern consisting of a stack of transformers as the encoder which is connected to the decoder using skip connections. 
A $1$D
sequence from the 3D input volume $x \in \mathcal{R}^{H\times W\times D\times C}$ with image dimension $(H,W,D)$ and C input channels is created by dividing $x$ into 
$N=(H\times W\times D)/P^{3}$ flattened  non-overlapping patches of size $P\times P\times P$ and denote this set as $x_v$. %T.
A linear layer projects the patches
onto a $K$ dimensional embedding space.%, .
To preserve spatial information a 1D learnable positional embedding $E_{pos} \in \mathcal{R}^{N\times K}$ is added to
the projected patch embedding $E \in \mathcal{R}^{P^{3}.C\times K}$ as $z_0=\left[\textbf{x}_v^1\textbf{E};\textbf{x}_v^2\textbf{E};\cdots\textbf{x}_v^N\textbf{E}\right] +\textbf{E}_{pos}$.
 Then multiple transformer blocks \cite{ViT} are used that have  multi-head self-attention (MSA) and multilayer perceptron (MLP) sublayers according to
 \begin{equation}
 \begin{split}
   z'_i & =MSA(\text{Norm}(z_{i-1}))+z_{i-1}, i=1\cdots L \\
   z_i & = MLP(\text{Norm}(z'_{i}))+z'_{i}, i=1\cdots L
   \end{split}
\end{equation} 
where $Norm()$ denotes layer normalization \cite{UNETR_1}, 
MLP has two linear layers with GELU activation functions,
$i$ denotes intermediate block and $L$ denotes transformer layers.
 $SA$ maps a query ($q$) and
the corresponding key ($k$) and value ($v$) representations
in a sequence $\textbf{z} \in \mathcal{R}^{N\times K}$. Attention weights ($A$) measure similarity between elements in $z$ and their key-value pairs according to $A=\text{Softmax} \left(\frac{\textbf{q}\textbf{k}^T}{\sqrt{K_h}} \right)$, where $K_h = K/n$ is a scaling factor. 
Thus,  $SA(\textbf{z})=\textbf{A}\textbf{v}$, where
$v$ denotes input sequence values, and MSA  output is:% defined as
\begin{equation}
  MSA= \left[SA_1(\textbf{z});SA_2(\textbf{z});\cdots;SA_n(\textbf{z}) \right]\textbf{W}_{msa},
\end{equation}
where $\textbf{W}_{msa} \in \mathcal{R}^{n.K_h\times K}$ represents the multi-headed trainable parameter weights.

\textbf{Self Supervised Learning:}
A pre-trained transformer network such as UNETR has the advantage of being trained on medical images. We take the UNETR network and finetune it in a self supervised manner using images from the different datasets that we use for super resolution. Self supervised finetuning of ViT has attracted a fair bit of attention of late with different approaches using contrastive learning \cite{ChenIccV21} and masked auto-encoding \cite{Cheniccv21_9,ViT}. We investigate both approaches and identify \cite{ViT} as more stable for our task. %, 
We remove the pre-trained prediction head and attach a zero-initialized $D \times K$ feedforward layer, where K is the number of downstream classes, and $D$ is the dimension of the flattened patches.
We define a pre-text task to identify the primary organ in the images, which is akin to a classification problem involving $K$ classes.

\subsection{Feature Disentanglement}

In order to separate the images into structure and texture components we train an autoencoder (AE)  shown in Figure~\ref{fig:workflow} (b). In a classic AE the encoder $E$ and generator $G$ form a mapping between image $x$ and latent code $z$ using an image reconstruction loss
\begin{equation}
    \mathcal{L}_{rec} (E,G)=\mathbb{E}_{x\sim X} \left[\left\|x-G(E(x)) \right\|_1  \right]
\end{equation}

To ensure that the generated image is realistic we have discriminator $D$ that calculates the adversarial loss for generator $G$ and encoder $E$ as:
\begin{equation}
    \mathcal{L}_{adv} (E,G,D)=\mathbb{E}_{x\sim X} \left[-\log(D(G(E(x)))) \right]
\end{equation}
    
As shown in Figure~\ref{fig:workflow} (b) we divide the latent code into two components - a texture component $z_t$ and a structural component $z_s$. Then amongst similar images $X^1,X^2$ from the same dataset in a minibatch we swap the two components and enforce the constraint that the resulting images be realistic, using the `swapped-GAN' loss \cite{SwapVAE} 
\begin{equation}
    \mathcal{L}_{swap} (E,G,D)=\mathbb{E}_{x^1,x^2\sim X,x^1\neq x^2} \left[-\log(D(G(z_s^1,z_t^2))) \right]
\end{equation}

Here $z_s^1,z_t^2$ are the first and second components of $E(x^1)$ and $E(x^2)$. The intuition is to combine the structure component of one image with the texture component of another image. The two images are not identical although they belong to the same dataset. As shown in Figure~\ref{fig:workflow} (b)  the shapes of $z_s$ and $z_t$ are asymmetric. $z_s$ is designed to be a tensor with spatial dimensions so it can learn the structural properties associated with spatial configurations, and $z_t$ is vector that encodes the texture information. 
At each training iteration we randomly sample two images $x^1$ and $x^2$, and enforce $\mathcal{L}_{rec},\mathcal{L}_{adv}$ for $x^1$, while applying $\mathcal{L}_{swap}$ to the combination of $x^1$ and $x^2$. 
The final loss function for \emph{feature disentanglement} is given in Eqn.~\ref{eq:loss2}, and more details are given in Appendix~\ref{app:feat}. 
\begin{equation}
    \mathcal{L}_{Disent} =\mathcal{L}_{Rec} +0.7\mathcal{L}_{Adv} + 0.7\mathcal{L}_{swap}
    \label{eq:loss2}
\end{equation}
We first train this disentanglement autoencoder that can extract the two separate features for a given input image (high or low resolution). The structure and texture features of the HR and LR images are used to train the super resolution network.

Since the HR and LR images are different versions of the same image swapping the structure  code $z_s^{LR}$ (or texture $z_t^{LR}$) of the LR image with that of the HR image $z_s^{HR}$ (or $z_t^{HR}$) should still generate an image that is close to the original. Patches of size $n\times n$ are extracted around the center of the LR image and corresponding patches of size $mn\times mn$ are extracted from the center of the HR image, $m$ being the upscaling factor. This ensures that the two patches show the same region of interest. Swapping  $z_t^{LR}$ with $z_t^{HR}$ and combining with $z_s^{LR}$ should produce an image very similar to the LR image. Similarly, $z_t^{LR}$ and $z_s^{HR}$ combine to give a fairly similar representation of the higher resolution image.

\textbf{Training The Super Resolution Network:}
We use two pre-trained networks - the ViT and the feature disentanglement network. Given the LR image $x$ and the intermediate  generated HR image $y$, we obtain their respective disentangled latent feature representations as $z^x_s,z^x_t$ and $z^y_s,z^y_t$. Thereafter we calculate the semantic similarity between them using the cosine similarity loss as 
\begin{equation}
\begin{split}
    \mathcal{L}_{str} & = 1-\langle z_s^{x},z_s^{y}\rangle \\
    \mathcal{L}_{tex} & = 1-\langle z_t^{x},z_t^{y}\rangle.
\end{split}
    \label{eq:disloss}
\end{equation}
where $\langle . \rangle$ denotes cosine similarity.
Additionally we also obtain the ViT based feature vectors of the HR ($f_{ViT}^{HR}$) and LR ($f_{ViT}^{LR}$) images from the ViT described previously and calculate their corresponding cosine similarity loss as 
\begin{equation}
    \mathcal{L}_{ViT} = 1-\langle f_{ViT}^{LR},f_{ViT}^{HR}\rangle
    \label{eq:vitloss}
\end{equation}
Once the above loss terms are obtained we train the whole super resolution network in an end to end manner using the following loss function.
Thus the final loss function is % 
\begin{equation}
    \mathcal{L_{SR}}(X,Y) = \mathcal{L}_{adv} +\lambda_1\mathcal{L}_{ViT}(X,Y) + \lambda_2\mathcal{L}_{str}(X,Y) + \lambda_3\mathcal{L}_{tex}(X,Y).
    \label{eq:loss1}
\end{equation}

\section{Experiments And Results}

\textbf{Dataset Description:}
We use two datasets for our experiments: 1) \textbf{fastMRI} \cite{Fastmri} - following \cite{FengMulti_21}, we filter
out $227$ and $24$ pairs of proton density (PD) and fat suppressed proton density weighted images (FS-PDWI) volumes for training and validation.
2) The \textbf{IXI dataset}: Three types of MR
images are included in the datasets (i.e., PD, T1, and T2)\footnote{http://brain-development.org/ixi-dataset/}.
Each of them has $500,70$, and $6$ MR volumes for training, testing, and validation respectively. Subvolumes of size $240\times240\times96$  are used and due to using 2D images we get $500\times96 = 48,000$ training samples.

\begin{table}[ht]
 \begin{center}
% \begin{tabular}{|c|c|c|c|c|c|c|}
\begin{adjustbox}{width=\textwidth}
% \begin{tabular}{|c|cc|cc|cc|cc|cc|cc|}
\begin{tabular}{|c|c|c|c|c|} 
\hline 
{} & \multicolumn{2}{|c|}{IXI - PD Images} & \multicolumn{2}{|c|}{IXI - T1 Images} \\ \hline
{} & {2$\times$} & {4$\times$} & {2$\times$} & {4$\times$} \\ \hline
{} & \textbf{PSNR/SSIM/NMSE} & \textbf{PSNR/SSIM/NMSE} & \textbf{PSNR/SSIM/NMSE} & \textbf{PSNR/SSIM/NMSE} \\ \hline
{Bicubic} & {30.4/0.9531/.042} & {29.13/0.8799/0.048} & {33.80/0.9525/0.030} & {28.28/0.8312/0.051} \\
\cite{FengTrans} & 31.7/ 0.892/ 0.035 & 29.5/ 0.870/ 0.033 & 30.7/ 0.883 /0.032 & 28.5/ 0.861/ 0.037  \\
\cite{ZhangCVPR21_8} & 38.96 / 0.9836/0.022 & 31.10 / 0.9181/0.030 & 37.12 / 0.9761/ 0.26 & 29.90 / 0.8796/ 0.034  \\
\cite{ZhangCVPR21_48} & 40.31 / 0.9870 / 0.021 & 32.73 / 0.9387/ 0.029 & 37.95 / 0.9795/ 0.028 & 31.05 / 0.9042/ 0.031  \\
\cite{ZhangCVPR21_49} & 41.28 / 0.9895 / 0.02 & 33.40 / 0.9486/ 0.027 & 
38.27 / 0.9810/ 0.025 & 31.23 / 0.9093/ 0.032 \\
\cite{ZhangCVPR21} & 41.66 /0.9902/0.019 & 33.97/0.9542/0.024 & 38.74/0.9824/0.021 & 32.03/0.9219/0.026  \\
{\cite{HuMIDL21}} & {42.9/0.9936/0.018} & {35.3/0.962/0.023} & {39.9/0.989/0.021} & {33.6/0.927/0.024} \\
\hline
{\textbf{Proposed}} & 44.3/0.9972/0.016 & 37.1/0.972/0.021 & 41.4/0.993/0.019 & 35.4/0.9386/0.022  \\ \hline
\multicolumn{5}{|c|}{\textbf{Ablation Studies}} \\ \hline
{$\mathcal{L}_{tex} + \mathcal{L}_{ViT} $} & 41.1/0.9826/0.021 & 35.3/0.958/0.025 & 39.5/0.983/0.021 & 33.1/0.924/0.024  \\ %\hline
{$\mathcal{L}_{str} + \mathcal{L}_{ViT} $} & 43.1/0.9902/0.018 & 35.8/0.963/0.023 & 40.5/0.986/0.021 & 34.5/0.9301/0.024  \\ %\hline
{$\mathcal{L}_{ViT} $} & 36.9/0.9745/0.027 & 34.2/0.943/0.027 & 37.2/0.962/0.025 & 31.3/0.903/0.028  \\ %\hline
{$\mathcal{L}_{tex} + \mathcal{L}_{str} $} & 37.6/0.9817/0.026 & 35.0/0.967/0.025 & 38.7/0.976/0.023 & 32.5/0.924/0.026 \\ \hline
\end{tabular}
\end{adjustbox}
\caption{Quantitative Results for IXI Dataset. Higher values of PSNR and SSIM, and lower value of NMSE indicate better results. }
\label{tab:IXI}
\end{center}
\end{table}

\subsection{Implementation Details}

\textbf{ViT Parameters}: For self supervised finetuning we use a batch size of $6$ and  cross entropy loss, the AdamW optimizer \cite{AdamW} with initial learning rate of $0.0001$ for $20,000$ iterations. For the specified batch size, the average training time was $10$ hours for $20,000$ iterations. %
\textbf{AE Network:} The encoder consists of 4 convolution blocks followed by max pooling after each step. The decoder is also symmetrically designed. $3\times3$ convolution filters are used and $64,64,32,32$ filters are used in each conv layer. The input to the AE is $256\times256$ and dimension of $z_{tex}$ is $256$, while $z_{str}$ is $64\times64$.

\textbf{Super Resolution Network:}
We train our model using Adam \cite{Adam}) with $\beta_1 = 0.9, \beta_2 = 0.999$, a batch size of $256$ and a weight decay of $0.1$, for $100$ epochs. We implement all models in PyTorch and train them using one NVIDIA Tesla V100 GPU with $32$GB of memory. $\lambda_1=\lambda_2=1$ and $\lambda_3=0.9$ (from Eqn.\ref{eq:loss1}).

\subsection{Quantitative Results}
For a given upscaling factor we first downsample the original image by that factor and recover the original size using different super resolution methods, and compare the performance using different metrics such as peak signal to noise ratio (PSNR), Structural Similarity Index Metric (SSIM), and Normalized Mean Square Error (NMSE).
Tables~\ref{tab:IXI},~\ref{tab:fastMRI} show the average values of different methods for  the IXI and fastMRI datasets at upscaling factors of $2\times$ and $4\times$.
Our method shows the best performance for both datasets and beats the next best method by a significant margin. While there is an expected noticeable performance drop for higher scaling factors, our method still outperforms other methods significantly. 
Our proposed method's advantage is the combination of CNN and ViT features that improve the image quality significantly.
Although image quality degrades at higher magnification factor, our method performs better than others due to its ability to leverage local and global information.

\textbf{Ablation Studies:}
Tables~\ref{tab:IXI},~\ref{tab:fastMRI} also show ablation study outcomes where different loss terms are excluded during training. Excluding the ViT features results in reduced performance. However it is still better than most other methods because of using feature disentanglement that leads to better super resolution based on texture and structure features. On the other hand excluding only one or more of structure and texture features leads to poor performance despite including ViT features. Thus we conclude that both global and local information is important for accurate super resolution.

\begin{table}[ht]
 \begin{center}
% \begin{tabular}{|c|c|c|c|c|c|c|}
\begin{adjustbox}{width=\textwidth}
% \begin{tabular}{|c|cc|cc|cc|cc|cc|cc|}
\begin{tabular}{|c|c|c|c|c|c|} 
\hline 
{} & \multicolumn{2}{|c|}{IXI-T2 Images} & {} & \multicolumn{2}{|c|}{Fast MRI} \\ \hline
{} & {2$\times$} & {4$\times$} & {} & {2$\times$} & {4$\times$}  \\ \hline
{} & \textbf{PSNR/SSIM/NMSE} & \textbf{PSNR/SSIM/NMSE} & {} & \textbf{PSNR/SSIM/NMSE} & \textbf{PSNR/SSIM/NMSE} \\ \hline
\cite{FengTrans}  & 30.2/0.891/0.034 & 28.4/0.878/0.033 & \cite{FengMulti_11} & 26.66/0.512/0.063 & 18.363/0.208/0.082  \\ 
\cite{ZhangCVPR21_8} &  37.32/0.9796/0.027 & 29.69/0.9052/0.031 & \cite{FengMulti_25} & 28.27/0.667/0.051 & 21.81/0.476/0.067 \\ 
\cite{ZhangCVPR21_48} & 38.75/0.9838/0.026 & 31.45/0.9324/0.029 & \cite{FengMulti_12} & 28.870/0.670/.048 & 23.255/0.507/0.062 \\
\cite{ZhangCVPR21_49} & 39.71/0.9863/0.027 & 32.05/0.9413/0.031 & \cite{VDSR} & 29.484/0.682/0.049 & 28.219/0.574/0.059 \\
\cite{ZhangCVPR21} & 40.30/0.9874/0.022 & 32.62 / 0.9472/0.029 & \cite{FengMulti} & 31.769/0.709/0.045 & 29.819/0.601/0.054 \\
{\cite{HuMIDL21}} & {41.9/0.991/0.020} & {34.2 / 0.951/0.027} & - & - & - \\ \hline
{\textbf{Proposed}} & 44.1/0.9953/0.017 & 35.4/0.959/0.024 & \textbf{Proposed} & 34.6/0.731/0.041 & 32.7/0.63/0.050 \\ \hline
\multicolumn{6}{|c|}{\textbf{Ablation Studies}} \\ \hline
{$\mathcal{L}_{tex} + \mathcal{L}_{ViT} $} & 40.8/0.977/0.021 & 33.6/0.941/0.027 & {$\mathcal{L}_{tex} + \mathcal{L}_{ViT} $} & 32.1/0.713/0.046 & 30.4/0.60/0.054  \\ %\hline
{$\mathcal{L}_{str} + \mathcal{L}_{ViT} $} & 43.0/0.9875/0.020 & 34.3/0.947/0.026 & {$\mathcal{L}_{str} + \mathcal{L}_{ViT} $} & 33.3/0.723/0.044 & 32.1/0.61/0.053  \\ %\hline
{$\mathcal{L}_{ViT} $} & 36.7/0.972/0.028 & 34.0/0.937/0.028 & {$\mathcal{L}_{ViT} $} & 30.1/0.694/0.048 & 28.9/0.59/0.056  \\ %\hline
{$\mathcal{L}_{tex} + \mathcal{L}_{str}$} & 36.9/0.980/0.026 & 34.6/0.948/0.026 & {$\mathcal{L}_{tex} + \mathcal{L}_{str}$} & 31.3/0.703/0.046 & 29.8/0.61/0.054 \\ \hline
\end{tabular}
\end{adjustbox}
\caption{Quantitative Results for IXI and fastMRI dataset super resolution output. Higher values of PSNR and SSIM, and lower value of NMSE indicate better results. }
\label{tab:fastMRI}
\end{center}
\end{table}

\subsection{Qualitative results:}
\begin{figure*}[t]
 \centering
\begin{tabular}{ccccc}
\includegraphics[height=2.5cm, width=1.6cm]{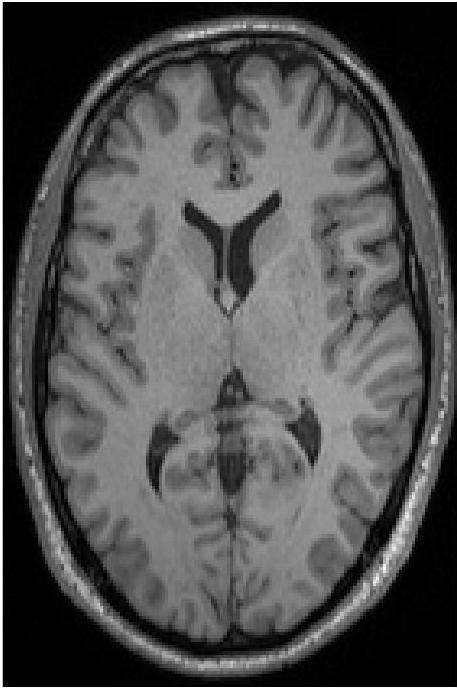} &
\includegraphics[height=2.5cm, width=1.6cm]{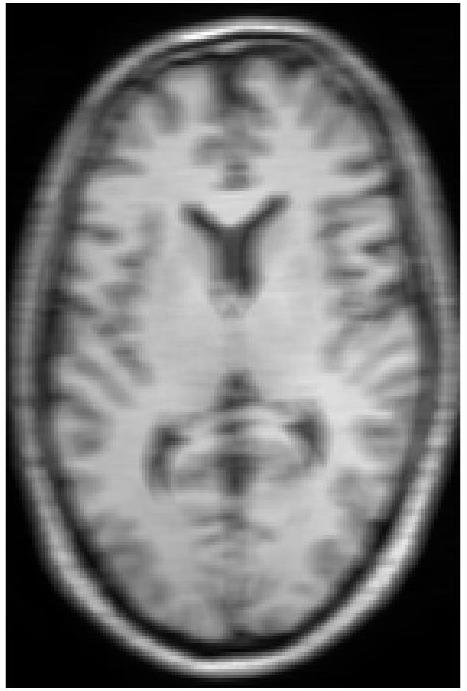} &
\includegraphics[height=2.5cm, width=1.6cm]{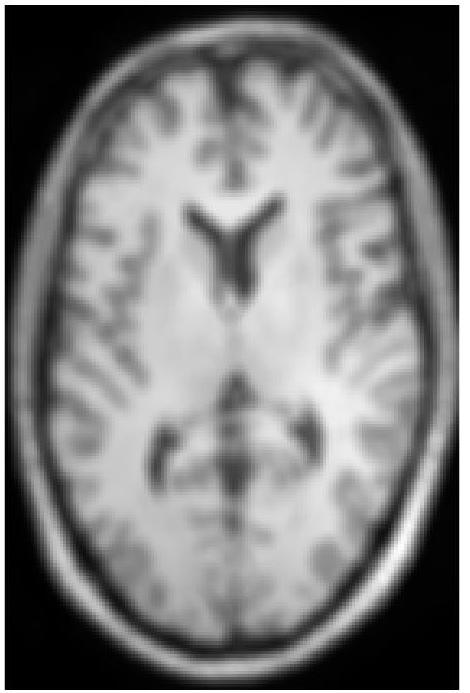} &
\includegraphics[height=2.5cm, width=1.6cm]{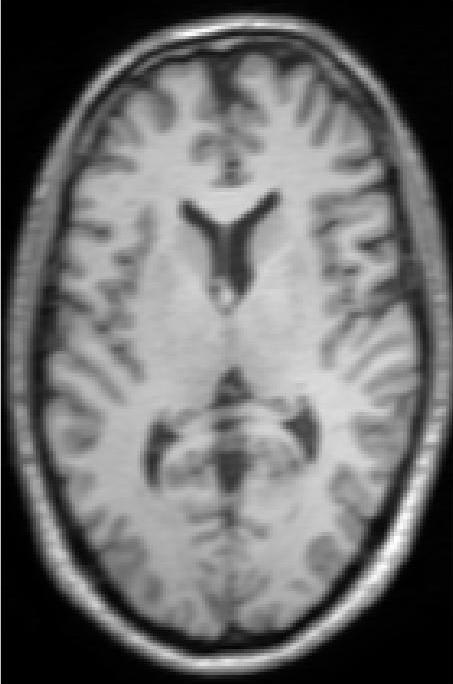} &
\includegraphics[height=2.5cm, width=1.6cm]{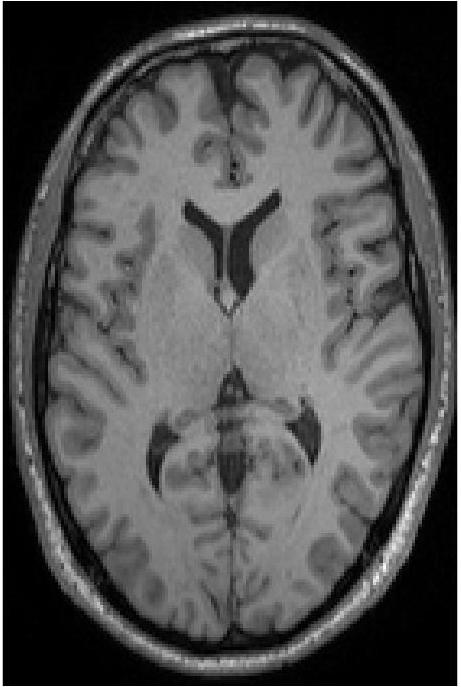}\\
&
\includegraphics[height=2.5cm, width=1.6cm]{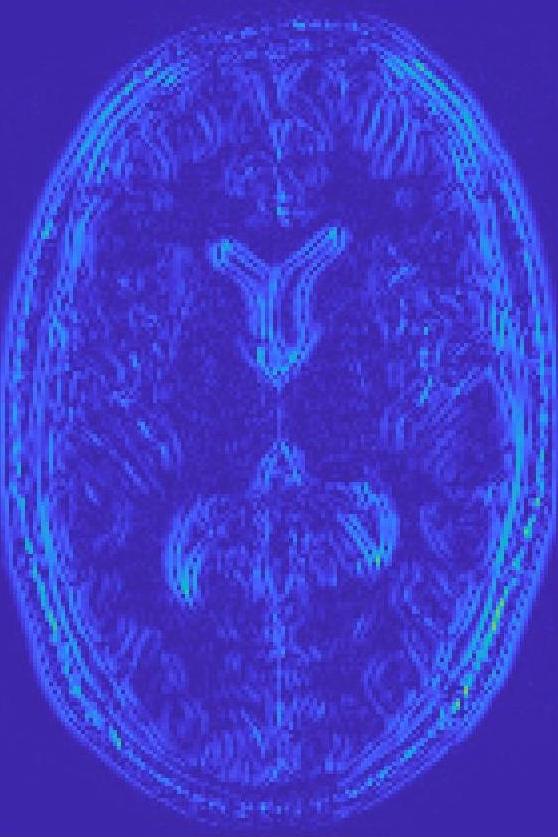} &
\includegraphics[height=2.5cm, width=1.6cm]{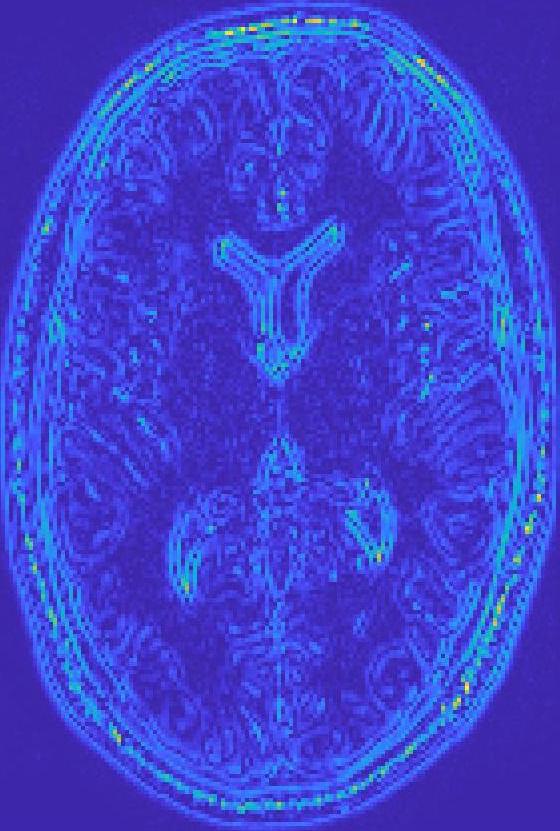} &
\includegraphics[height=2.5cm, width=1.6cm]{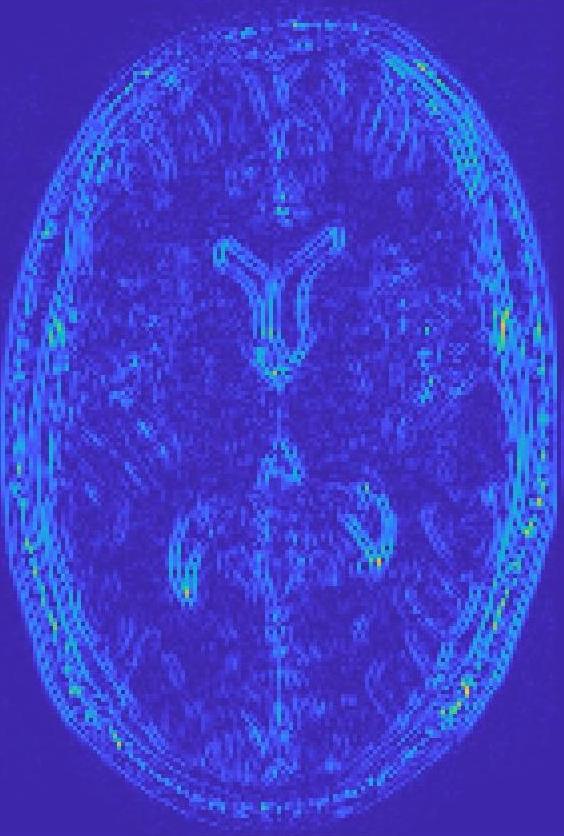} &
\includegraphics[height=2.5cm, width=1.9cm]{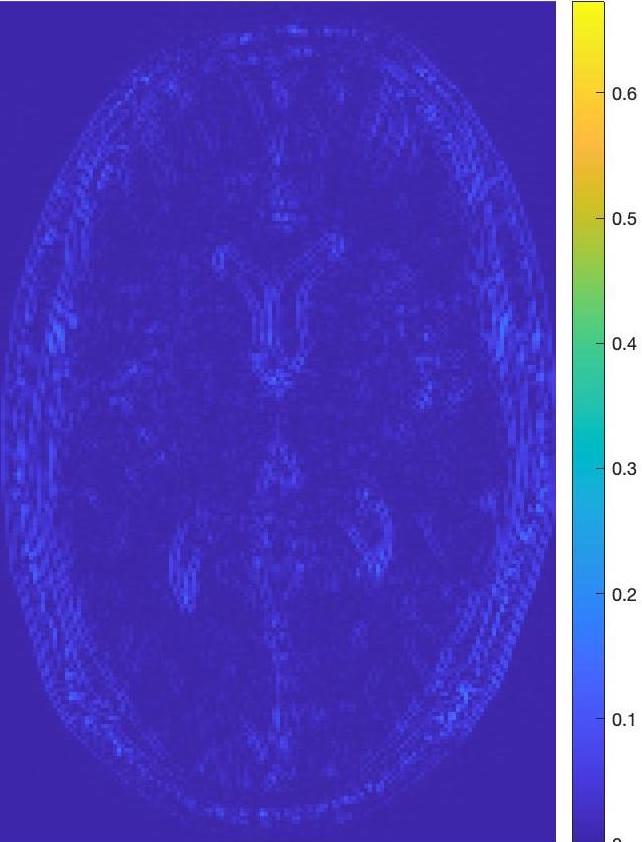}\\
(a) & (b) & (c) & (d) & (e)\\
\end{tabular}
\caption{Visualization of superresolution results at $2\times$ factor for the IXI dataset. The top row dhows the original image and the super resolved images and the bottom row shows the corresponding difference images.(a) Original image; Superesolved images obtianed using: b) \cite{ZhangCVPR21}; (c) \cite{ZhangCVPR21_49}; (d) \cite{ZhangCVPR21_48}; (e) Our proposed method.}
\label{fig:ISR}
\end{figure*}

In Figure~\ref{fig:ISR} we show visualization results where the recovered images and their corresponding difference image with the original image is shown. Our method shows a very accurate reconstruction   with minimal regions in the error map, while the recovered images from other methods are blurred and of poor quality. These results demonstrate the effectiveness of our approach.

\section{Conclusion}

We proposed a novel method for MR image super resolution by combining CNNs and Vision transformers. ViTs provide more global context features while CNNs provide discriminative local information. We achieve feature disentanglement using swapped auto encoders to obtain texture and structure features. We enforce constraints that the original and super resolved images should have similar semantic information by minimizing the cosine loss of the respective structure and texture features, as well as minimizing the difference between the respective ViT features. Experimental results show our method outperforms state of the art techniques on benchmark public datasets, and ablation studies demonstrate the importance of our proposed loss terms.

\appendix

\section{Feature Disentanglement}
\label{app:feat}

Similar to a classic autoencoder, the encoder $E$ produces a latent code $z \sim Z$ for image $x\sim X$. The $G$ reconstructs the original image from $z$ using an image reconstruction loss that is defined as:
\begin{equation}
    \mathcal{L}_{Rec}(E,G)=\mathbb{E}_{x\sim X}\left[\left\|x-G(E(x)) \right\| \right]
\end{equation}
Additionally, the generated image should be realistic as determined by the Discriminator $D$ and is enforced using the adversarial loss defined as:
\begin{equation}
    \mathcal{L}_{Adv}(E,G,D)=\mathbb{E}_{x\sim X}\left[-\log(D(G(E(x)))) \right]
\end{equation}

Furthermore, as part of our objective to achieve feature disentanglement we decompose the latent code $z$ into two components $[z_{str},z_{tex}]$ corresponding to the structure and texture components. We enforce that swapping these components of the latent code with those from other images still produces realistic images. This is achieved by using a modified version of the adversarial loss, which we term as the swapped GAN loss, and is defined as :
\begin{equation}
    \mathcal{L}_{swap} (E,G,D)=\mathbb{E}_{x^1,x^2\sim X,x^1\neq x^2} \left[-\log(D(G(z_{tex}^1,z_{str}^2))) \right]
\end{equation}
Here $z_{tex}^1,z_{str}^2$ are the first and second components of images $X^1,X^2$'s latent representations, and $X^1,X^2$ from the same dataset in a minibatch. The component $z_{str}$ is a tensor with spatial dimensions, while $z_{tex}$ is a vector that encode structure and
texture information. $\mathcal{L}_{Rec}$ and $\mathcal{L}_{Adv}$, are applied to image $X^1$ while $\mathcal{L}_{swap}$ is applied to the latent components from $X^1,X^2$.The final loss function for feature disentanglement is defined as 
\begin{equation}
    \mathcal{L}_{Disent} =\mathcal{L}_{Rec} +0.7\mathcal{L}_{Adv} + 0.7\mathcal{L}_{swap}
\end{equation}

\section{UNETR Architecture }
\label{app:UNETR}

We use the  ViT from UNETR  \cite{UNETR} pre-trained for MR image segmentation and describe its architecture below. %hanges that improve performance over UNETR
UNETR uses the contracting-expanding pattern consisting of a stack of transformers as the encoder which is connected to the decoder using skip connections. 
A 1D sequence of 3D input volume $x \in \mathcal{R}^{H\times W\times D\times C}$ with image dimension $(H,W,D)$ and C input channels is created by dividing it into
flattened uniform non-overlapping patches $x_v \in \mathbb{R}^{N\times(P^3.C}$ 
where $P\times P\times P$ denotes the resolution of each patch and
$N=(H\times W\times D)/P^{3}$ is the length of the sequence.

A linear layer projects the patches
onto a $K$ dimensional embedding space which remains
constant throughout the transformer layers.
To preserve spatial information a 1D learnable positional embedding $E_{pos} \in \mathcal{R}^{N\times K}$ is added to
the projected patch embedding $E \in \mathcal{R}^{P^{3}.C\times K}$ according to 
% 
%
% according to the following 
\begin{equation}
    z_0=\left[\textbf{x}_v^1\textbf{E};\textbf{x}_v^2\textbf{E};\cdots\textbf{x}_v^N\textbf{E}\right] +\textbf{E}_{pos}
\end{equation} 
 Then multiple transformer blocks \cite{ViT} are used that have  multi-head self-attention (MSA) and multilayer perceptron (MLP) sublayers according to
\begin{equation}
  z'_i=MSA(\text{Norm}(z_{i-1}))+z_{i-1}, i=1\cdots L
\end{equation} 
\begin{equation}
  z_i= MLP(\text{Norm}(z'_{i}))+z'_{i}, i=1\cdots L
\end{equation} 
% %
where $Norm()$ denotes layer normalization \cite{UNETR_1}, 
MLP has two linear layers with GELU activation functions,
$i$ denotes intermediate block and $L$ denotes transformer layers.
A MSA sublayer comprises of n parallel self-attention
(SA) heads. Specifically, the SA block, is a parameterized
function that  maps a query ($q$) and
the corresponding key ($k$) and value ($v$) representations
in a sequence $\textbf{z} \in \mathcal{R}^{N\times K}$. Attention weights ($A$) measure similarity between elements in $z$ and their key-value pairs according to % 
\begin{equation}
  A=\text{Softmax} \left(\frac{\textbf{q}\textbf{k}^T}{\sqrt{K_h}} \right).
\end{equation} 
 $K_h = K/n$ is a scaling factor for  maintaining the
number of parameters to a constant value with different
values of the key \textbf{k}. 
Using the computed attention weights,
the output of SA for values v in the sequence z is computed as
\begin{equation}
   SA(\textbf{z})=\textbf{A}\textbf{v},
\end{equation} 
$v$ denotes input sequence values, and MSA  output is:% defined as
\begin{equation}
  MSA= \left[SA_1(\textbf{z});SA_2(\textbf{z});\cdots;SA_n(\textbf{z}) \right]\textbf{W}_{msa},
\end{equation}
where $\textbf{W}_{msa} \in \mathcal{R}^{n.K_h\times K}$ represents the multi-headed trainable parameter weights.

At the encoder bottleneck (i.e. output of transformer’s
last layer),  a deconvolutional layer is applied to the
transformed feature map to increase its resolution by a factor
of 2.  The resized feature map is concatenated with the
feature map of the previous transformer output and
fed  into consecutive $3\times3\times3$ convolutional layers, whose
output is upsampled using a deconvolutional layer. This
process is repeated for all the other subsequent layers up
to the original input resolution where the final output is fed
into a $1\times1\times1$ convolutional layer with a softmax activation
function to generate voxel-wise semantic predictions.

\subsection{Loss Function}

The loss function is a combination of soft dice loss 
and cross-entropy loss, and it can be computed in a voxel-wise
manner according to
\begin{equation}
    \mathcal{L}(G,Y)=1-\frac{2}{J} \sum_{j=1}^J \frac{\sum_{i=1}^I G_{i,j}Y_{i,j} }{\sum_{i=1}^I G^2_{i,j} + \sum_{i=1}^I Y^2_{i,j}} - \frac{1}{I} \sum_{i=1}^I\sum_{j=1}^J G_{i,j} \log Y_{i,j}
\end{equation}
where $I$ is the number of voxels; $J$ is the number of classes;
$Y_{i,j}$ and $G_{i,j}$ denote the probability output and one-hot
encoded ground truth for class $j$ at voxel $i$, respectively. For a detailed explanation of all terms we urge the reader to refer \cite{UNETR}.

The UNETR was implemented by the authors  in PyTorch and MONAI and  trained using a NVIDIA DGX-1 server. All models were trained with the batch size of 6, using the AdamW optimizer \cite{AdamW}
with initial learning rate of 0.0001 for 20,000 iterations.
For the specified batch size, the average training time was 10 hours for 20,000 iterations.
The transformer-based
encoder follows the ViT-B16 \cite{ViT} architecture with L=12
layers, an embedding size of K=768. The patch resolution 
was $16\times16\times16$. For inference  a sliding window was used
 with an overlap portion of $0.5$ between the neighboring
patches.
The authors did not use any pre-trained weights for the transformer
backbone (e.g. ViT on ImageNet) since it did not demonstrate
any performance improvements for the medical images.

% \subsection{Self Supervised Learning}

\section{Additional Visual Results}

\begin{figure*}[t]
 \centering
\begin{tabular}{c}
\includegraphics[ width=13.6cm,keepaspectratio]{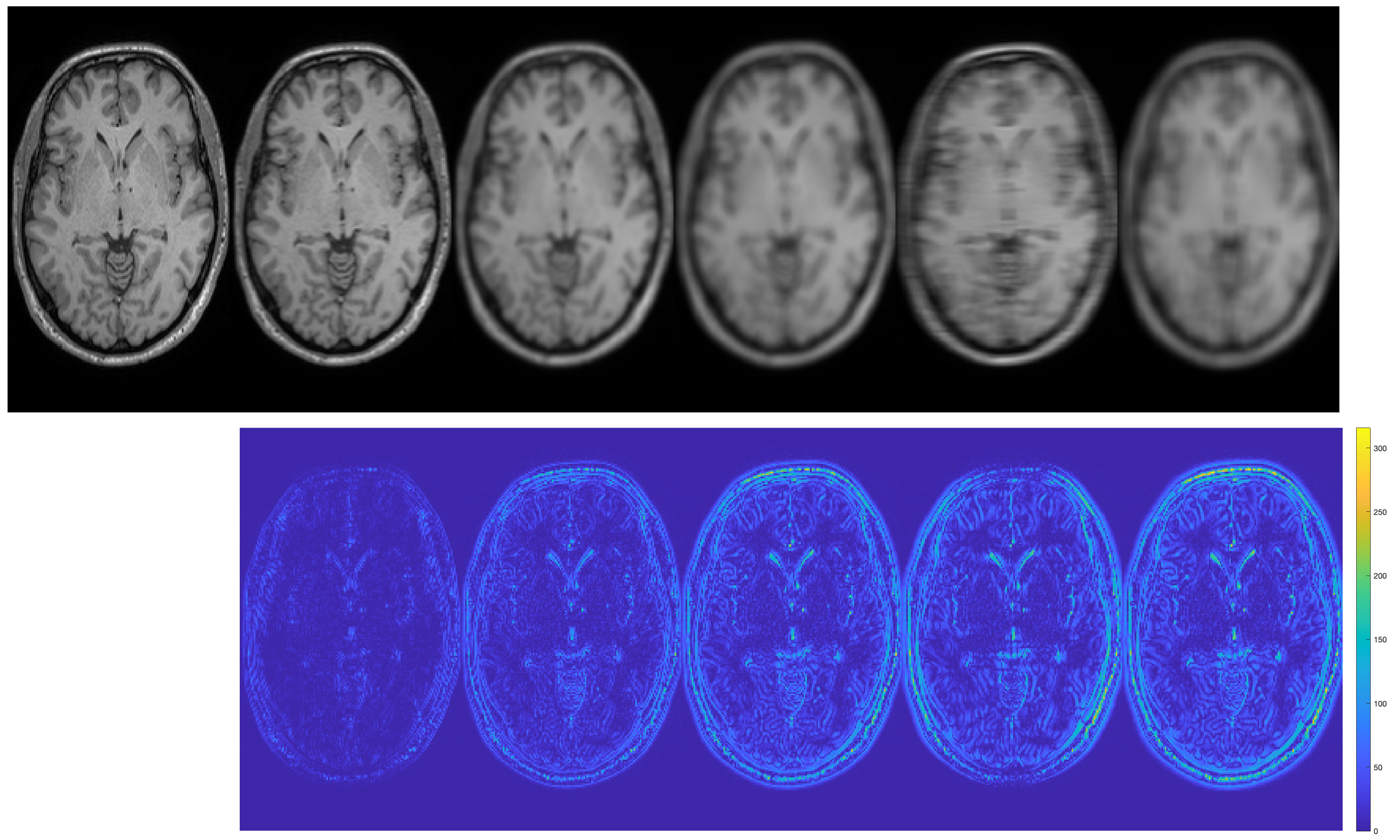} \\
\includegraphics[ width=13.6cm,keepaspectratio]{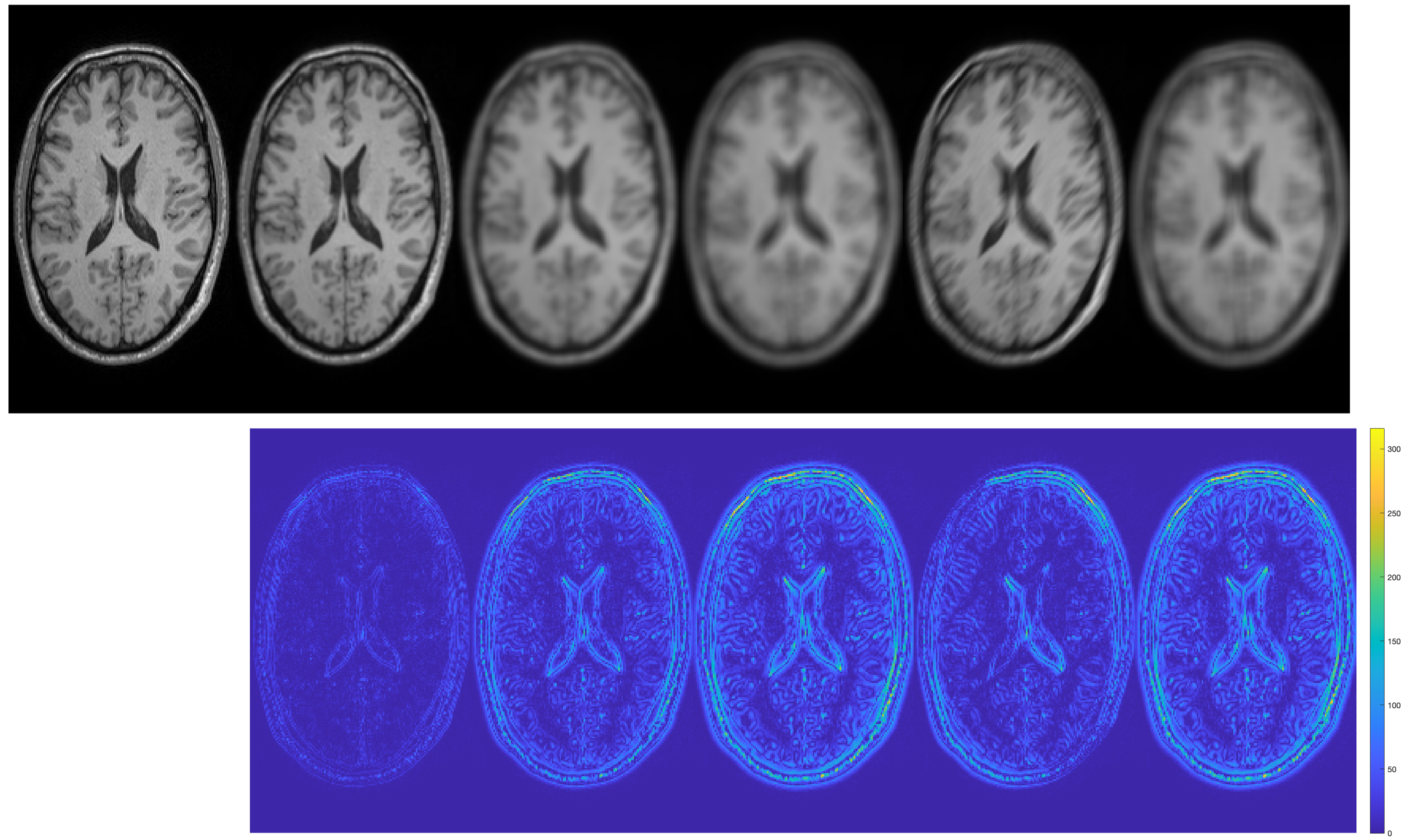} \\
\text{For IXI Image Dataset at 2x superresolution} \\
\end{tabular}
% \begin{tabular}{c}
% \includegraphics[height=3.3cm, width=15cm]{Diff_combined.jpg}  \\
% \end{tabular}
\caption{Visualization of superresolution. For each figure the top row is the original image followed by the difference image inthe bottom row. Column 1- original image; Reconstructed Image using: Column 2- Our Proposed method; Column 3 - $\mathcal{L}_{str} + \mathcal{L}_{ViT} $; Column 4 - $\mathcal{L}_{tex} + \mathcal{L}_{ViT} $; Column 5 - $\mathcal{L}_{tex} + \mathcal{L}_{str}$; Column 6 - $\mathcal{L}_{ViT} $.}
\label{fig:ISR2}
\end{figure*}

\begin{figure*}[t]
 \centering
\begin{tabular}{c}
\includegraphics[ width=13.6cm,keepaspectratio]{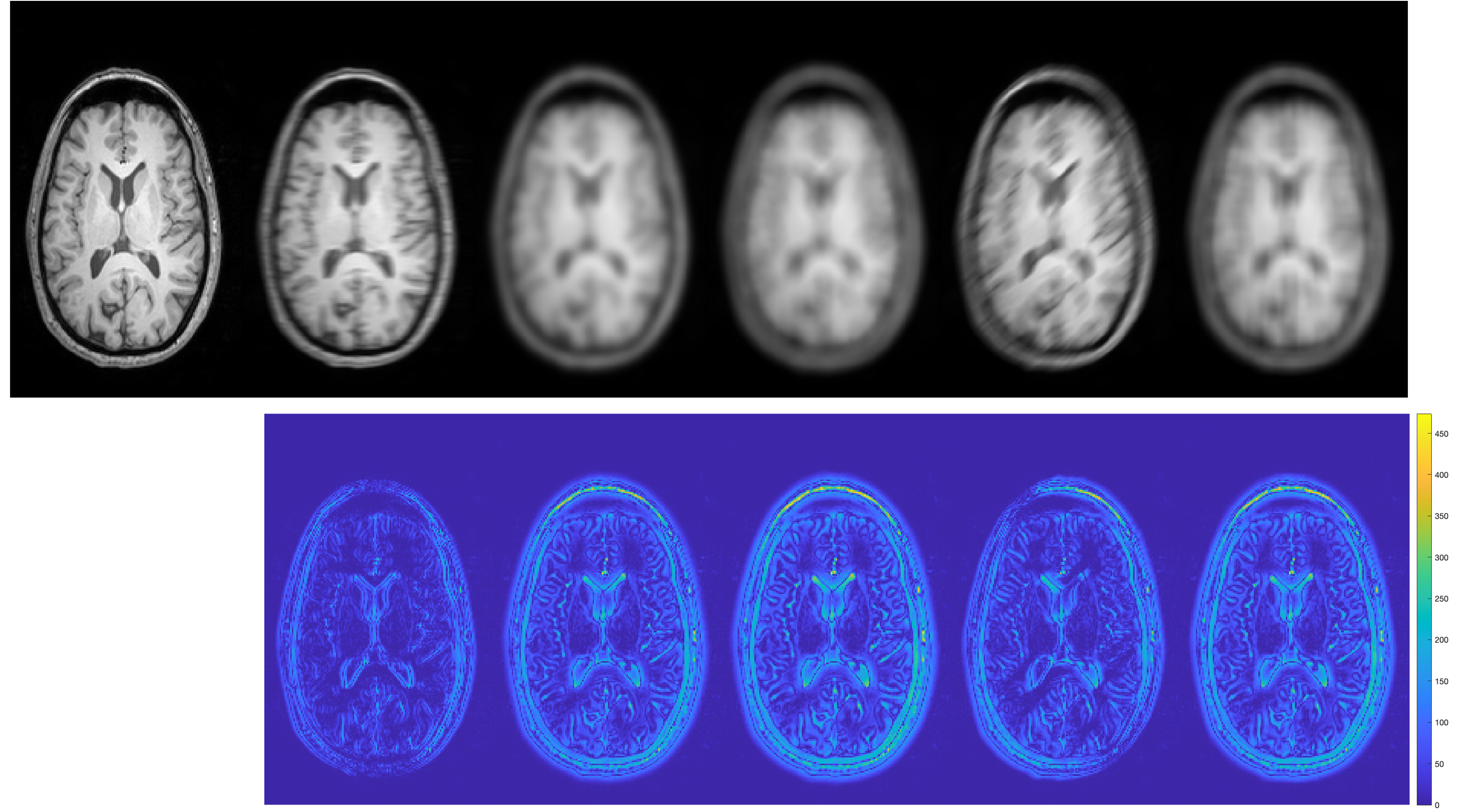} \\
\includegraphics[ width=13.6cm,keepaspectratio]{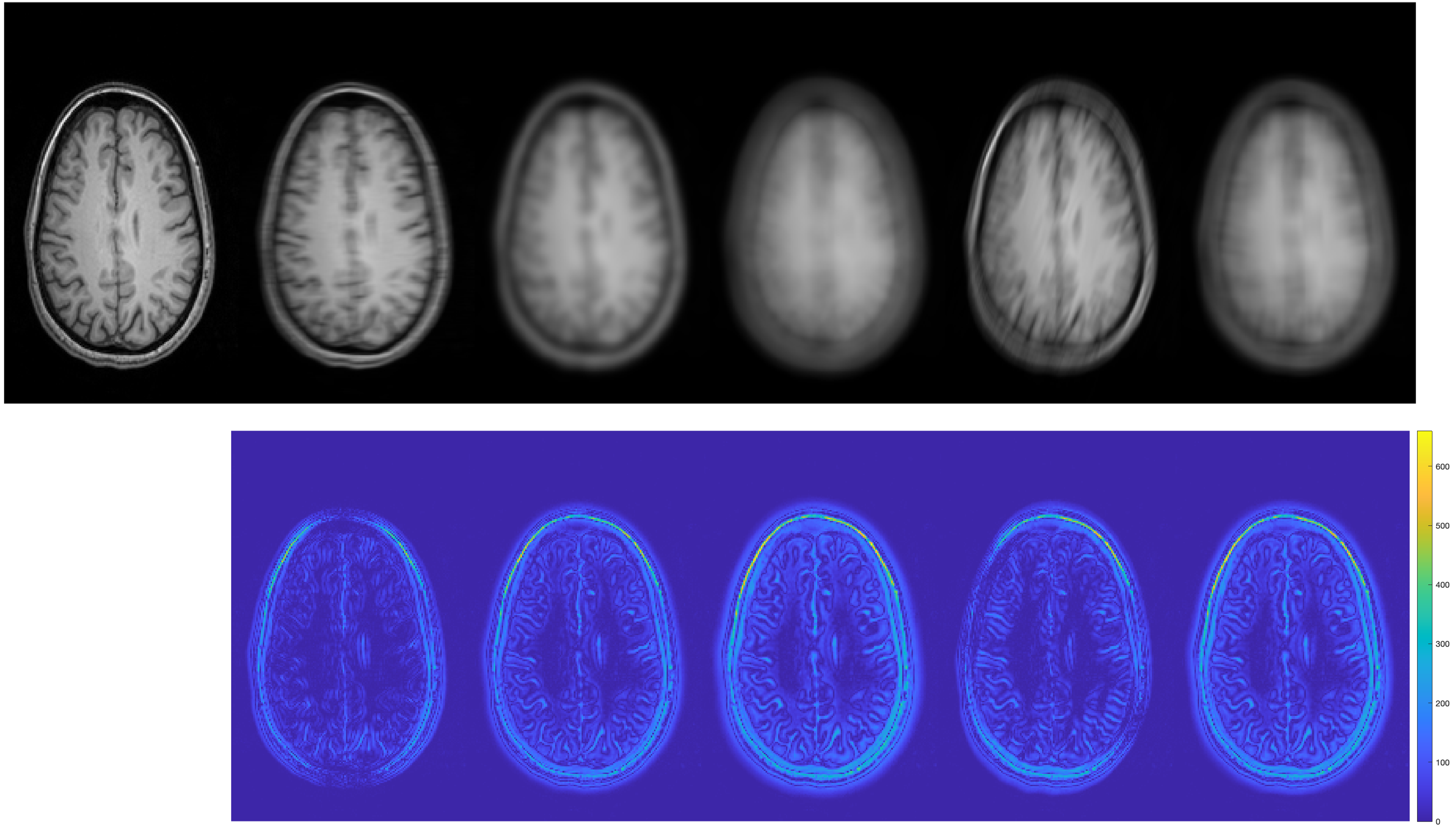} \\
\text{For IXI Image Dataset at 4x superresolution} \\
\end{tabular}
% \begin{tabular}{c}
% \includegraphics[height=3.3cm, width=15cm]{Diff_combined.jpg}  \\
% \end{tabular}
\caption{Visualization of superresolution. For each figure the top row is the original image followed by the difference image inthe bottom row. Column 1- original image; Reconstructed Image using: Column 2- Our Proposed method; Column 3 - $\mathcal{L}_{str} + \mathcal{L}_{ViT} $; Column 4 - $\mathcal{L}_{tex} + \mathcal{L}_{ViT} $; Column 5 - $\mathcal{L}_{tex} + \mathcal{L}_{str}$; Column 6 - $\mathcal{L}_{ViT} $.}
\label{fig:ISR3}
\end{figure*}

\begin{figure*}[t]
 \centering
\begin{tabular}{c}
\includegraphics[ width=13.6cm,keepaspectratio]{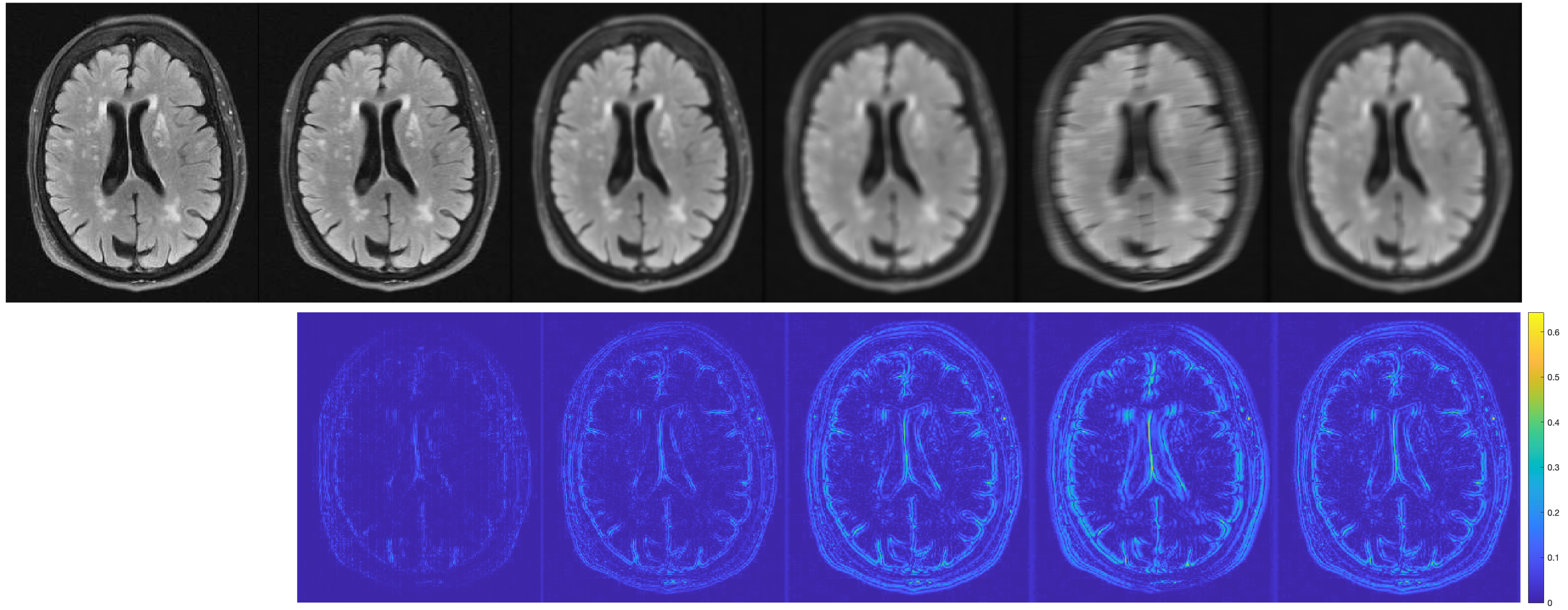} \\
\text{For FastMRI Image Dataset at 2x superresolution} \\
\includegraphics[ width=13.6cm,keepaspectratio]{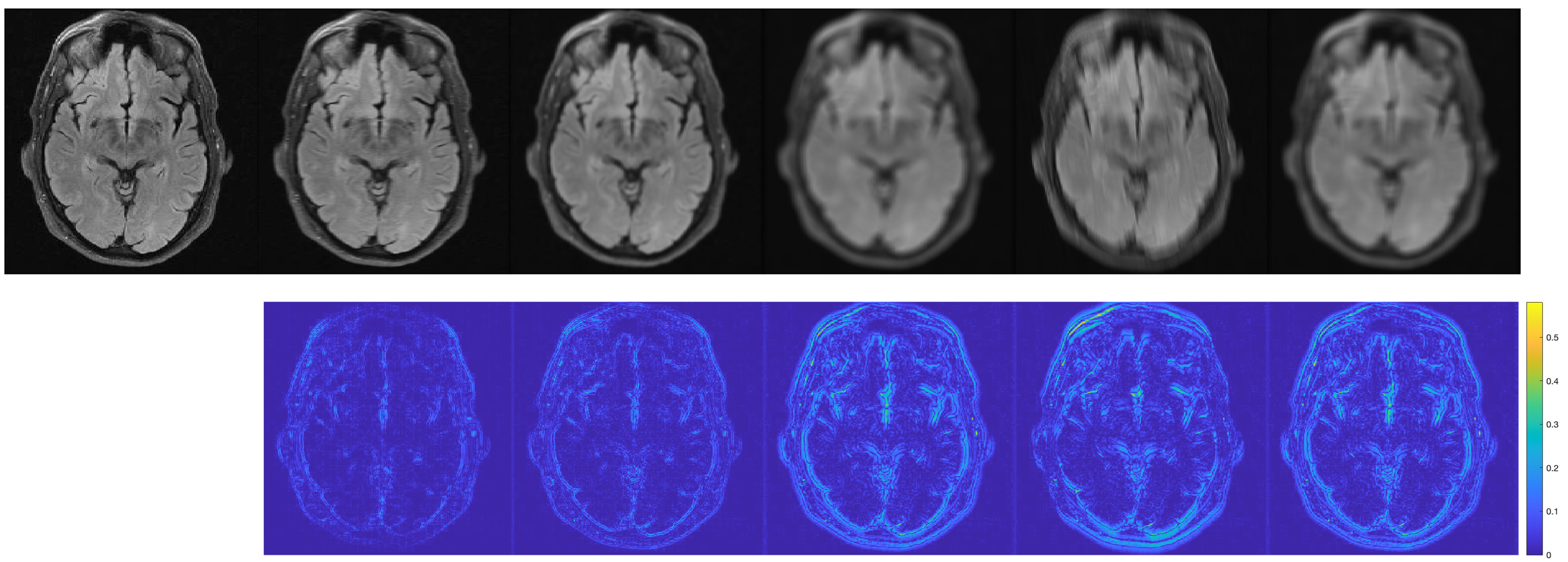} \\
\text{For FastMRI Image Dataset at 4x superresolution} \\
\end{tabular}
% \begin{tabular}{c}
% \includegraphics[height=3.3cm, width=15cm]{Diff_combined.jpg}  \\
% \end{tabular}
\caption{Visualization of superresolution results. For each figure the top row is the original image followed by the difference image inthe bottom row. Column 1- original image; Reconstructed Image using: Column 2- Our Proposed method; Column 3 - $\mathcal{L}_{str} + \mathcal{L}_{ViT} $; Column 4 - $\mathcal{L}_{tex} + \mathcal{L}_{ViT} $; Column 5 - $\mathcal{L}_{tex} + \mathcal{L}_{str}$; Column 6 - $\mathcal{L}_{ViT} $.}
\label{fig:ISR4}
\end{figure*}

In this section we show additional visual results (Figures~\ref{fig:ISR2},\ref{fig:ISR3},\ref{fig:ISR4}) from the IXI and Fast MRI dataset at different super resolution factors for the different ablation settings. The figures show the original image and the reconstructed image along with the difference image. They clearly illustrate the important contribution of each of the loss terms, and the adverse impact on super resolution if we exclude different terms.

\section{Computation Time}

The original UNETR model has $92.58$ Million parameters, and our finetuned model has similar number of parameters at $93.4$ Million. The training time on a NVIDIA Tesla V100 GPU was $10$ hours for $20,000$ iterations for the finetuning stage. The feature disentanglement network took $18$ hours to train for $100$ epochs.
For the actual super resolution step, it took us $14$ hours to train for $80$ epochs. Note that feature disentanglement and ViT finetuning wer pre-trained and while training the super resolution network we only extracted features from them

The original UNETR model's inference time was $12.08$s. Feature extraction from the finetuned UNETR model took $1.3$s, while the disentangled feature extraction took $0.05$ seconds per image. For the actual super resolution at inference stage it took $1.2$ seconds for $2$x upsampling for a $512\times512$ image

\section{Architecture of Super Resolution Network}

\begin{figure*}[t]
 \centering
\begin{tabular}{c}
\includegraphics[ width=13.6cm,keepaspectratio]{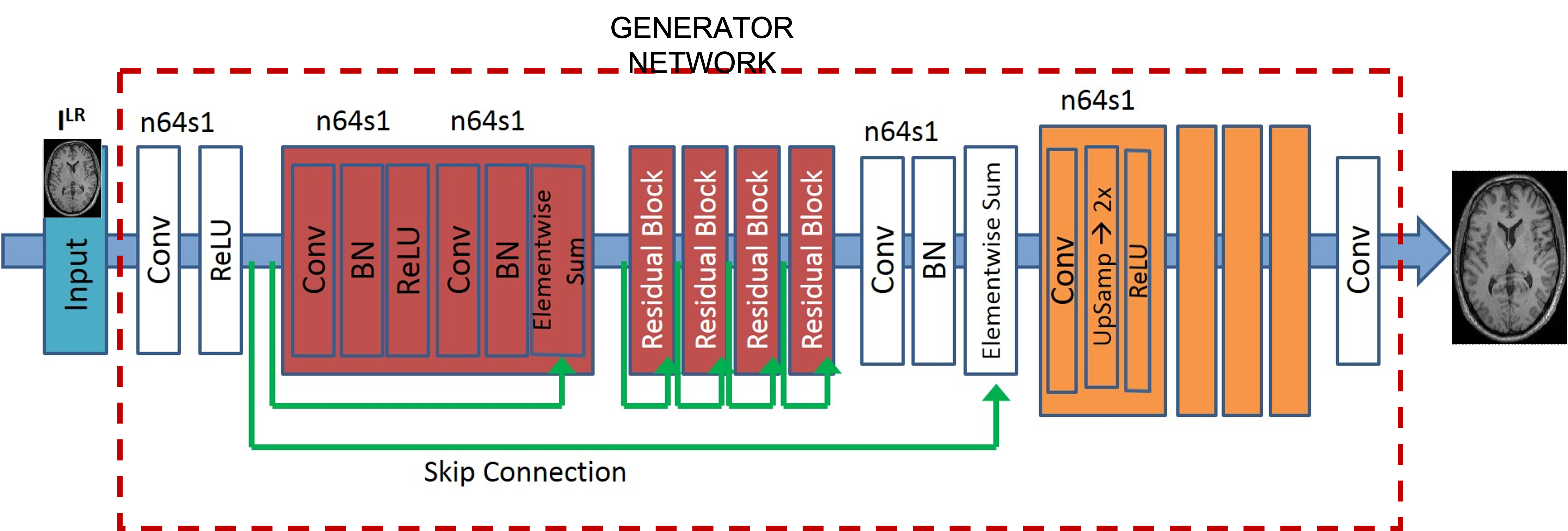} \\
(a) \\
\includegraphics[ width=13.6cm,keepaspectratio]{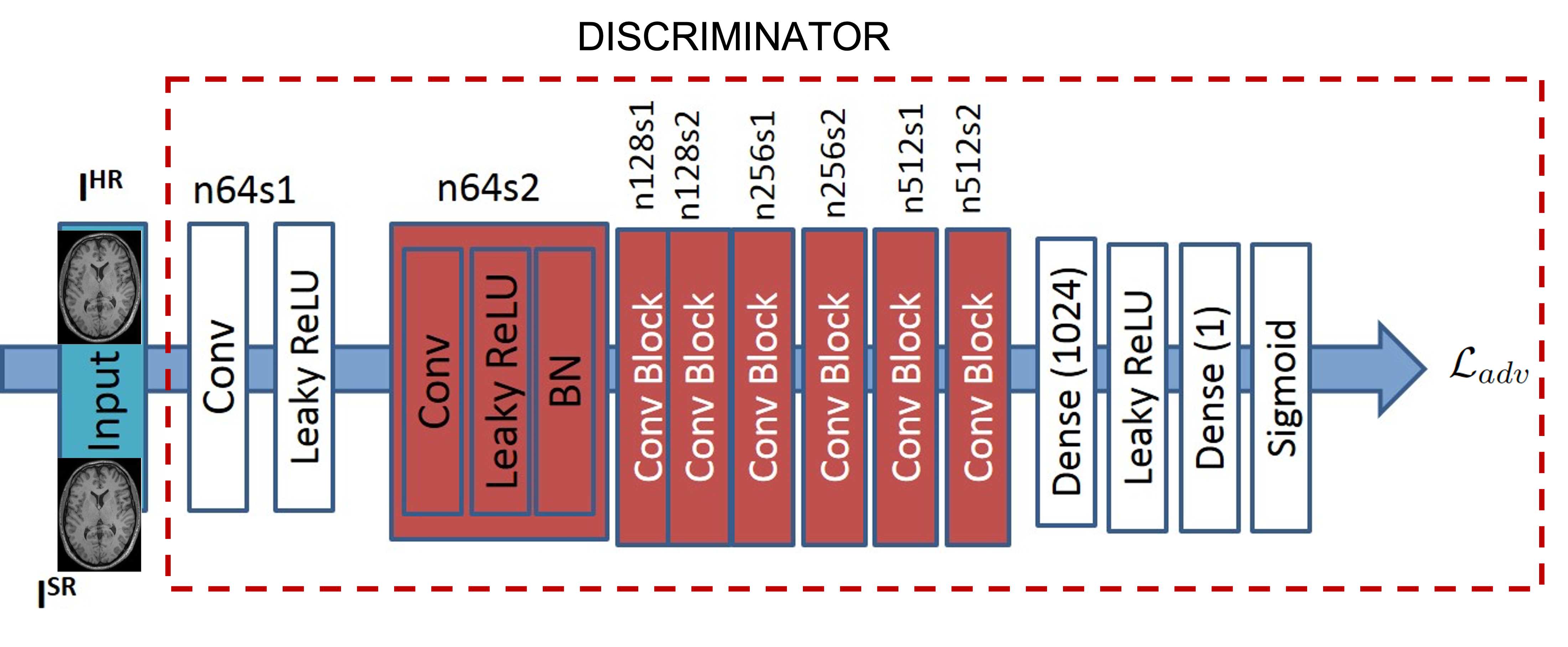} \\
 (b) \\
\end{tabular}
\caption{(a) Generator Network; (b) Discriminator network. $n64s1$ denotes $64$ feature maps (n) and stride (s) $1$ for each convolutional layer..}
\label{fig:gen}
\end{figure*}

Figure~\ref{fig:gen} shows the detailed architecture of the super resolution network's generator and discriminator components. In the generator (Figure~\ref{fig:gen} (a)) the input low resolution image $I^{LR}$ is passed through a convolution block followed by ReLU activation. The output is passed through a residual block with skip connections. Each block has convolutional layers with $3\times3$ filters and $64$ feature maps,  followed by batch normalization and ReLU activation. This output is subsequently passed through multiple residual blocks. Their output is passed through a series of upsampling stages, where each stage doubles the input image size. The output is passed through a convolution stage to get the super resolved image $I^{SR}$. Depending upon the desired scaling, the number of upsampling stages can be changed.  The discriminator outputs the $\mathcal{L}_{adv}$ in Eqn~\ref{eq:loss1}, and is defined as:
\begin{equation}
    \mathcal{L}_{adv,SR} (E,G,D)=\mathbb{E}_{lr\sim LR} \left[-\log(D(G(E(lr)))) \right]
\end{equation}
where $LR$ is the set of low resolution images and $G(E(lr))$ is the super resolved high resolution image.
The other two loss terms, $\mathcal{L}_{ViT},\mathcal{L}_{tex},\mathcal{L}_{str}$, have been defined before.

\section{Loss Plots}

\begin{figure*}[t]
 \centering
\begin{tabular}{c}
\includegraphics[ width=13.6cm,keepaspectratio]{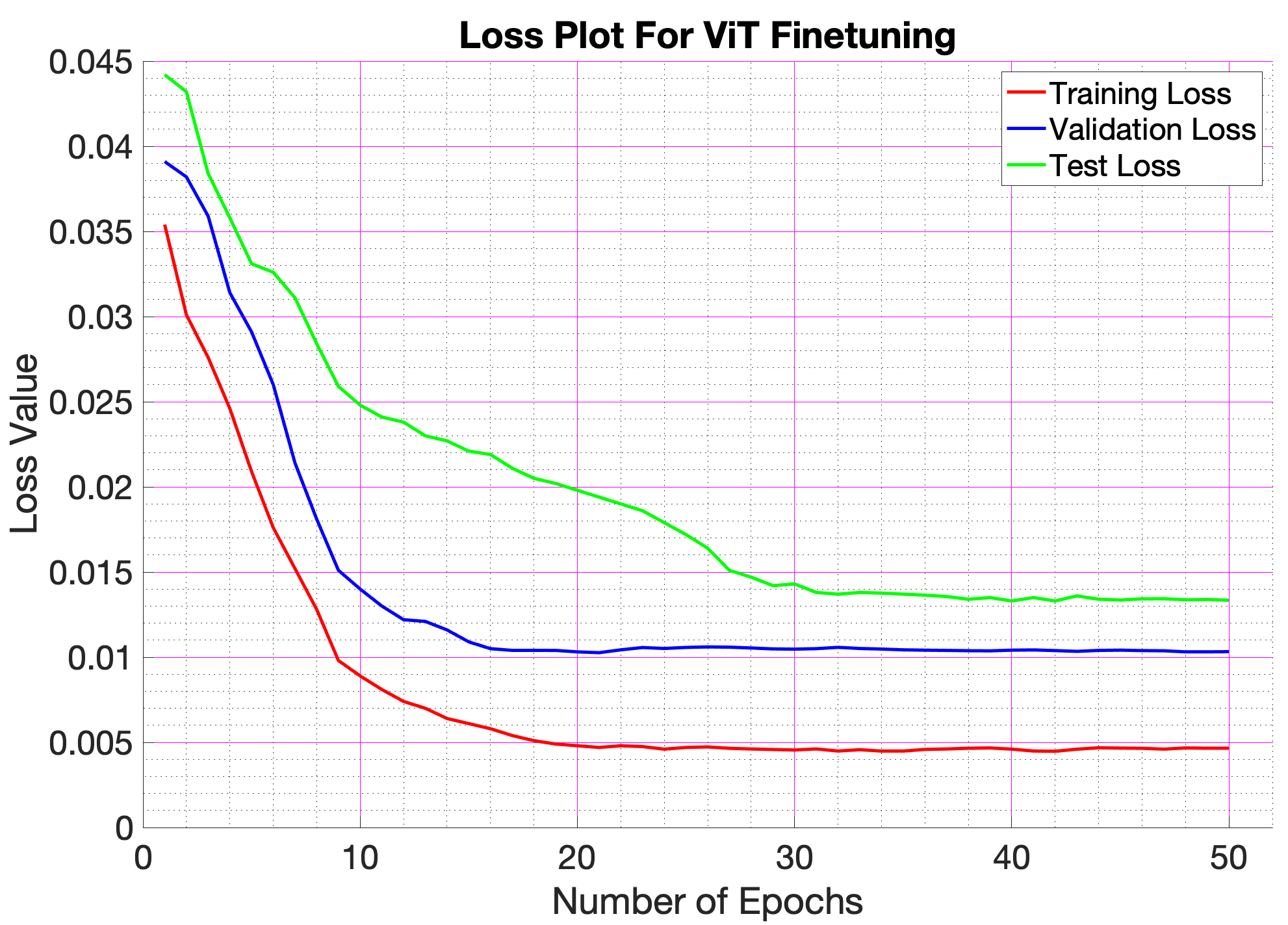} \\
(a) \\
\includegraphics[ width=13.6cm,keepaspectratio]{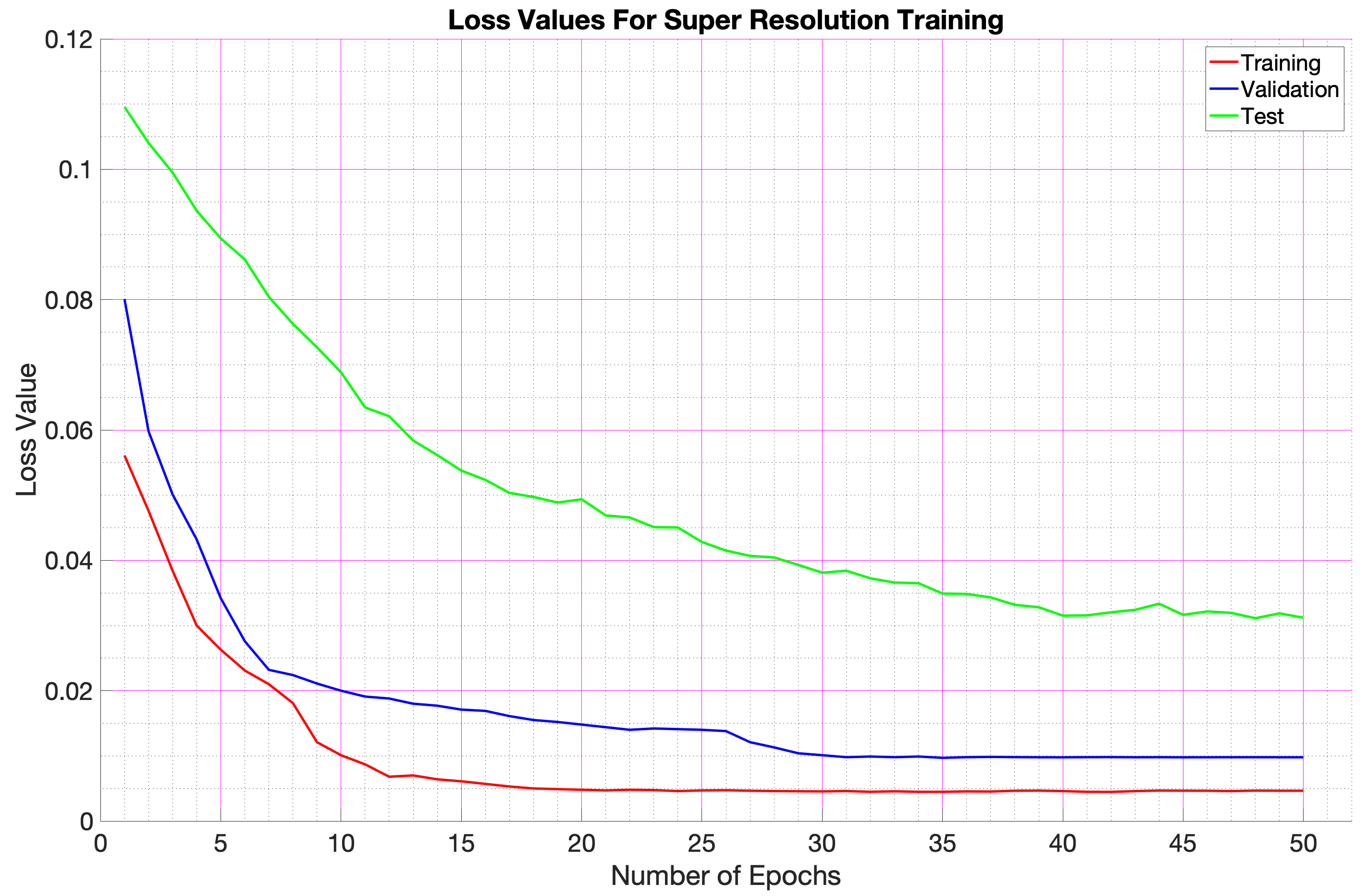} \\
 (b) \\
\end{tabular}
\caption{Loss plots for (a) UNETR Fine tuning using the IXI dataset; (b) Image super-resolution training for IXI dataset.}
\label{fig:loss}
\end{figure*}

In figure~\ref{fig:loss} we show the loss plots for training, validation and test data splits ont he IXI brain image dataset. We see that the training error decreases gradually , which is also observable for the validation and test errors, although their magnitudes are higher than the training error. The plots show that there is minimal chance of overfitting of the models and the results are not biased.

\section{Comparison With \cite{FengTrans}}

The results of \cite{FengTrans} come up as worse than bicubic interpolation on the IXI dataset. This is surprising considering that they use a task transformer network. In our re-implementation we report better results than those reported on the paper \cite{FengTrans} since we devote significant bit of time in finetuning the parameters. Our experiments show that by removing the task transformer component the performance degrades but is still better than the numbers in \cite{FengTrans}. While it is difficult to ascertain the reason behind their low performance, a possible reason could be the architecture of the task transformer network. This requires further investigation and is beyond the scope of our current work.

\bibliographystyle{splncs04}
\bibliography{mahapatra}

\begin{thebibliography}{100}
\providecommand{\url}[1]{\texttt{#1}}
\providecommand{\urlprefix}{URL }
\providecommand{\doi}[1]{https://doi.org/#1}

\bibitem{Pat15}
Antony, B., Sedai, S., Mahapatra, D., Garnavi, R.: Real-time passive monitoring
  and assessment of pediatric eye health. In: US Patent App. 16/178,757 (2020)

\bibitem{ZhangCVPR21_1}
Ba, J., Mnih, V., Kavukcuoglu, K.: Multiple object recognition with visual
  attention (2015)

\bibitem{UNETR_1}
Ba, J.L., Kiros, J.R., Hinton, G.E.: Layer normalization (2016)

\bibitem{Pat7}
Bastide, P., Kiral-Kornek, I., Mahapatra, D., Saha, S., Vishwanath, A.,
  Cavallar, S.V.: Machine learned optimizing of health activity for
  participants during meeting times. In: US Patent App. 15/426,634 (2018)

\bibitem{Pat5}
Bastide, P., Kiral-Kornek, I., Mahapatra, D., Saha, S., Vishwanath, A.,
  Cavallar, S.V.: Visual health maintenance and improvement. In: US Patent
  9,993,385 (2018)

\bibitem{Health_p}
Bastide, P., Kiral-Kornek, I., Mahapatra, D., Saha, S., Vishwanath, A.,
  Cavallar, S.V.: Crowdsourcing health improvements routes. In: US Patent App.
  15/611,519 (2019)

\bibitem{Mahapatra_CVIU2019}
Bozorgtabar, B., Mahapatra, D., von Teng, H., Pollinger, A., Ebner, L., Thiran,
  J.P., Reyes, M.: Informative sample generation using class aware generative
  adversarial networks for classification of chest xrays. Computer Vision and
  Image Understanding  \textbf{184},  57--65 (2019)

\bibitem{CVIU_Ar}
Bozorgtabar, B., Mahapatra, D., von Teng, H., Pollinger, A., Ebner, L., Thiran,
  J.P., Reyes, M.: Informative sample generation using class aware generative
  adversarial networks for classification of chest xrays. In: arXiv preprint
  arXiv:1904.10781 (2019)

\bibitem{Behzad_PR2020}
Bozorgtabar, B., Mahapatra, D., Thiran, J.P.: Exprada: Adversarial domain
  adaptation for facial expression analysis. In Press Pattern Recognition
  \textbf{100},  15--28 (2020)

\bibitem{Behzad_MICCAI20}
Bozorgtabar, B., Mahapatra, D., Thiran, J.P., Shao, L.: {SALAD}:
  Self-supervised aggregation learning for anomaly detection on x-rays. In: In
  Proc. MICCAI. pp. 468--478 (2020)

\bibitem{Salad_AR}
Bozorgtabar, B., Mahapatra, D., Vray, G., Thiran, J.P.: Anomaly detection on
  x-rays using self-supervised aggregation learning. In: arXiv preprint
  arXiv:2010.09856 (2020)

\bibitem{Frontiers2020}
Bozorgtabar, B., Mahapatra, D., Zlobec, I., Rau, T., Thiran, J.: Computational
  pathology. Frontiers in Medicine  \textbf{7} (2020)

\bibitem{Bozorgtabar_ICCV19}
Bozorgtabar, B., Rad, M.S., Mahapatra, D., Thiran, J.P.: Syndemo: Synergistic
  deep feature alignment for joint learning of depth and ego-motion. In: In
  Proc. IEEE ICCV (2019)

\bibitem{UNETR_5}
Carion, N., Massa, F., Synnaeve, G., Usunier, N., Kirillov, A., Zagoruyko, S.:
  End-to-end object detection with transformers. In: Vedaldi, A., Bischof, H.,
  Brox, T., Frahm, J.M. (eds.) Computer Vision -- ECCV 2020. pp. 213--229.
  Springer International Publishing, Cham (2020)

\bibitem{Unetr7}
Chen, J., Lu, Y., Yu, Q., Luo, X., Adeli, E., Wang, Y., Lu, L., Yuille, A.L.,
  Zhou, Y.: Transunet: Transformers make strong encoders for medical image
  segmentation (2021)

\bibitem{Cheniccv21_9}
Chen, M., Radford, A., Child, R., Wu, J., Jun, H., Luan, D., Sutskever, I.:
  Generative pretraining from pixels. In: III, H.D., Singh, A. (eds.)
  Proceedings of the 37th International Conference on Machine Learning.
  Proceedings of Machine Learning Research, vol.~119, pp. 1691--1703. PMLR
  (13--18 Jul 2020), \url{https://proceedings.mlr.press/v119/chen20s.html}

\bibitem{ChenIccV21}
Chen, X., Xie, S., He, K.: An empirical study of training self-supervised
  vision transformers (2021)

\bibitem{ZhangCVPR21_5}
Chen, Y., Shi, F., Christodoulou, A.G., Xie, Y., Zhou, Z., Li, D.: Efficient
  and accurate mri super-resolution using a generative adversarial network and
  3d multi-level densely connected network. In: MICCAI. pp. 91--99 (2018)

\bibitem{ZhangCVPR21_6}
Chen, Y., Xie, Y., Zhou, Z., Shi, F., Christodoulou, A.G., Li, D.: Brain {MRI}
  super resolution using 3d deep densely connected neural networks. CoRR
  \textbf{abs/1801.02728} (2018), \url{http://arxiv.org/abs/1801.02728}

\bibitem{DarTMI}
Dar, S.U., Yurt, M., Karacan, L., Erdem, A., Erdem, E., Çukur, T.: Image
  synthesis in multi-contrast mri with conditional generative adversarial
  networks. IEEE Transactions on Medical Imaging  \textbf{38}(10),  2375--2388
  (2019). \doi{10.1109/TMI.2019.2901750}

\bibitem{Souryaisbi22}
Das, S.D., Dutta, S., Shah, N.A., Mahapatra, D., Ge, Z.: Anomaly detection in
  retinal images using multi-scale deep feature sparse coding. In: 2022 IEEE
  19th International Symposium on Biomedical Imaging (ISBI). pp.~1--5 (2022).
  \doi{10.1109/ISBI52829.2022.9761713}

\bibitem{SouryaISBI_Ar}
Das, S.D., Dutta, S., Shah, N.A., Mahapatra, D., Ge, Z.: Anomaly detection in
  retinal images using multi-scale deep feature sparse coding. In: arXiv
  preprint arXiv:2201.11506 (2022)

\bibitem{Devika_IEEE}
Devika, K., Mahapatra, D., Subramanian, R., Oruganti, V.R.M.: Outlier-based
  autism detection using longitudinal structural mri. IEEE Access  \textbf{10},
   27794--27808 (2022). \doi{10.1109/ACCESS.2022.3157613}

\bibitem{DevikaAccess_Ar}
Devika, K., Mahapatra, D., Subramanian, R., Oruganti, V.R.M.: Outlier-based
  autism detection using longitudinal structural mri. In: arXiv preprint
  arXiv:2202.09988 (2022)

\bibitem{ZhangCVPR21_7}
Dong, C., Loy, C.C., He, K., Tang, X.: Learning a deep convolutional network
  for image super-resolution. In: ECCV. pp. 184--199 (2014)

\bibitem{ZhangCVPR21_8}
Dong, C., Loy, C.C., He, K., Tang, X.: Image super-resolution using deep
  convolutional networks. IEEE Transactions on Pattern Analysis and Machine
  Intelligence  \textbf{38},  295--307 (2016)

\bibitem{ViT}
Dosovitskiy, A., Beyer, L., Kolesnikov, A., Weissenborn, D., Zhai, X.,
  Unterthiner, T., Dehghani, M., Minderer, M., Heigold, G., Gelly, S.,
  Uszkoreit, J., Houlsby, N.: An image is worth 16x16 words: Transformers for
  image recognition at scale. CoRR  \textbf{abs/2010.11929} (2020),
  \url{https://arxiv.org/abs/2010.11929}

\bibitem{Esser}
Esser, P., Rombach, R., Ommer, B.: Taming transformers for high-resolution
  image synthesis (2020)

\bibitem{FengMulti}
Feng, C.M., Fu, H., Yuan, S., Xu, Y.: Multi-contrast mri super-resolution via a
  multi-stage integration network. In: de~Bruijne, M., Cattin, P.C., Cotin, S.,
  Padoy, N., Speidel, S., Zheng, Y., Essert, C. (eds.) Medical Image Computing
  and Computer Assisted Intervention -- MICCAI 2021. pp. 140--149. Springer
  International Publishing, Cham (2021)

\bibitem{FengTrans}
Feng, C.M., Yan, Y., Fu, H., Chen, L., Xu, Y.: Task transformer network for
  joint mri reconstruction and super-resolution. In: de~Bruijne, M., Cattin,
  P.C., Cotin, S., Padoy, N., Speidel, S., Zheng, Y., Essert, C. (eds.) Medical
  Image Computing and Computer Assisted Intervention -- MICCAI 2021. pp.
  307--317. Springer International Publishing, Cham (2021)

\bibitem{Pat14}
Garnavi, R., Mahapatra, D., Roy, P., Tennakoon, R.: System and method to teach
  and evaluate image grading performance using prior learned expert knowledge
  base. In: US Patent App. 10,657,838 (2020)

\bibitem{ZGe_MTA2019}
Ge, Z., Mahapatra, D., Chang, X., Chen, Z., Chi, L., Lu, H.: Improving
  multi-label chest x-ray disease diagnosis by exploiting disease and health
  labels dependencies. In press Multimedia Tools and Application pp. 1--14
  (2019)

\bibitem{Xr_Ar}
Ge, Z., Mahapatra, D., Sedai, S., Garnavi, R., Chakravorty, R.: Chest x-rays
  classification: A multi-label and fine-grained problem. In: arXiv preprint
  arXiv:1807.07247 (2018)

\bibitem{UNETR}
Hatamizadeh, A., Tang, Y., Nath, V., Yang, D., Myronenko, A., Landman, B.,
  Roth, H., Xu, D.: Unetr: Transformers for 3d medical image segmentation
  (2021)

\bibitem{Pat17}
Hoog, J.D., Mahapatra, D., Garnavi, R., Jalali, F.: Personalized monitoring of
  injury rehabilitation through mobile device imaging. In: US Patent App.
  16/589,046 (2021)

\bibitem{ZhangCVPR21_13}
Hu, J., Shen, L., Sun, G.: Squeeze-and-excitation networks. CoRR
  \textbf{abs/1709.01507} (2017), \url{http://arxiv.org/abs/1709.01507}

\bibitem{HuMIDL21}
Hu, X., Yan, Y., Ren, W., Li, H., Bayat, A., Zhao, Y., Menze, B.: Feedback
  graph attention convolutional network for {MR} images enhancement by
  exploring self-similarity features. In: Heinrich, M., Dou, Q., de~Bruijne,
  M., Lellmann, J., Schläfer, A., Ernst, F. (eds.) Proceedings of the Fourth
  Conference on Medical Imaging with Deep Learning. Proceedings of Machine
  Learning Research, vol.~143, pp. 327--337. PMLR (07--09 Jul 2021),
  \url{https://proceedings.mlr.press/v143/hu21a.html}

\bibitem{ZhangCVPR21_14}
Hu, Y., Li, J., Huang, Y., Gao, X.: Channel-wise and spatial feature modulation
  network for single image super-resolution. CoRR  \textbf{abs/1809.11130}
  (2018), \url{http://arxiv.org/abs/1809.11130}

\bibitem{ZhangCVPR21_15}
Hui, Z., Wang, X., Gao, X.: Fast and accurate single image super-resolution via
  information distillation network. In: CVPR. pp. 723--731 (2018)

\bibitem{ZhangCVPR21_16}
Iqbal, Z., Nguyen, D., Hangel, G., Motyka, S., Bogner, W., Jiang, S.:
  Super-resolution 1h magnetic resonance spectroscopic imaging utilizing deep
  learning. Frontiers in Oncology  \textbf{9} (2019)

\bibitem{ZhangCVPR21_17}
Jain, S., Sima, D.M., Nezhad, F.S., Williams, S., Van~Huffel, S., Maes, F.,
  Smeets, D.: Patch based super-resolution of mr spectroscopic images. In: 2016
  IEEE 13th International Symposium on Biomedical Imaging (ISBI). pp. 452--456
  (2016)

\bibitem{Lie_AR2}
Ju, L., Wang, X., Wang, L., Liu, T., Zhao, X., Drummond, T., Mahapatra, D., Ge,
  Z.: Relational subsets knowledge distillation for long-tailed retinal
  diseases recognition. In: arXiv preprint arXiv:2104.11057 (2021)

\bibitem{Lie_AR}
Ju, L., Wang, X., Wang, L., Mahapatra, D., Zhao, X., Harandi, M., Drummond, T.,
  Liu, T., Ge, Z.: Improving medical image classification with label noise
  using dual-uncertainty estimation. In: arXiv preprint arXiv:2103.00528 (2020)

\bibitem{LieTMI_2022}
Ju, L., Wang, X., Wang, L., Mahapatra, D., Zhao, X., Zhou, Q., Liu, T., Ge, Z.:
  Improving medical images classification with label noise using
  dual-uncertainty estimation. IEEE Transactions on Medical Imaging pp.~1--1
  (2022). \doi{10.1109/TMI.2022.3141425}

\bibitem{JuJbhi2020}
Ju, L., Wang, X., Zhao, X., Lu, H., Mahapatra, D., Bonnington, P., Ge, Z.:
  Synergic adversarial label learning for grading retinal diseases via
  knowledge distillation and multi-task learning. IEEE JBHI  \textbf{100},
  1--14 (2020)

\bibitem{LieMiccai21}
Ju, L., Wang, X., Zhao, X., Lu, H., Mahapatra, D., Ge, Z.: Relational subsets
  knowledge distillation for long-tailed retinal diseases recognition. In: In
  MICCAI 2021. pp. 1--11 (2021)

\bibitem{VDSR}
Kim, J., Lee, J.K., Lee, K.M.: Accurate image super-resolution using very deep
  convolutional networks. In: The IEEE Conference on Computer Vision and
  Pattern Recognition (CVPR Oral) (June 2016)

\bibitem{Adam}
Kingma, D., Ba, J.: Adam: A method for stochastic optimization. In: arXiv
  preprint arXiv:1412.6980, (2014)

\bibitem{Kuanar_AR1}
Kuanar, S., Athitsos, V., Mahapatra, D., Rajan, A.: Multi-scale deep learning
  architecture for nucleus detection in renal cell carcinoma microscopy image.
  In: arXiv preprint arXiv:2104.13557 (2021)

\bibitem{Kuanar_ICIP19}
Kuanar, S., Athitsos, V., Mahapatra, D., Rao, K., Akhtar, Z., Dasgupta, D.: Low
  dose abdominal ct image reconstruction: An unsupervised learning based
  approach. In: In Proc. IEEE ICIP. pp. 1351--1355 (2019)

\bibitem{Kuanar_AR2}
Kuanar, S., Mahapatra, D., Athitsos, V., Rao, K.: Gated fusion network for sao
  filter and inter frame prediction in versatile video coding. In: arXiv
  preprint arXiv:2105.12229 (2021)

\bibitem{Haze_Ar}
Kuanar, S., Rao, K., Mahapatra, D., Bilas, M.: Night time haze and glow removal
  using deep dilated convolutional network. In: arXiv preprint arXiv:1902.00855
  (2019)

\bibitem{KuanarVC}
Kuanar, S., Mahapatra, D., Bilas, M., Rao, K.: Multi-path dilated convolution
  network for haze and glow removal in night time images. The Visual Computer
  \textbf{38}(3),  1121--1134 (2022)

\bibitem{KuangAMM14}
Kuang, H., Guthier, B., Saini, M., Mahapatra, D., Saddik, A.E.: A real-time
  smart assistant for video surveillance through handheld devices. In: In Proc:
  ACM Intl. Conf. Multimedia. pp. 917--920 (2014)

\bibitem{LiTMI_2015}
Li, Z., Mahapatra, D., J.Tielbeek, Stoker, J., van Vliet, L., Vos, F.: Image
  registration based on autocorrelation of local structure. IEEE Trans. Med.
  Imaging  \textbf{35}(1),  63--75 (2016)

\bibitem{FengMulti_11}
Lim, B., Son, S., Kim, H., Nah, S., Lee, K.M.: Enhanced deep residual networks
  for single image super-resolution (2017)

\bibitem{UNETR_30}
Liu, Z., Lin, Y., Cao, Y., Hu, H., Wei, Y., Zhang, Z., Lin, S., Guo, B.: Swin
  transformer: Hierarchical vision transformer using shifted windows (2021)

\bibitem{AdamW}
Loshchilov, I., Hutter, F.: Decoupled weight decay regularization (2019)

\bibitem{FengMulti_12}
Lyu, Q., Shan, H., Steber, C., Helis, C., Whitlow, C., Chan, M., Wang, G.:
  Multi-contrast super-resolution mri through a progressive network. IEEE
  Transactions on Medical Imaging  \textbf{39}(9),  2738--2749 (2020).
  \doi{10.1109/TMI.2020.2974858}

\bibitem{MahapatraRegBook}
Mahapatra, D.: Elastic registration of cardiac perfusion images using saliency
  information. Sequence and Genome Analysis – Methods and Applications pp.
  351--364 (2011)

\bibitem{MahapatraMiccaiIAHBD11}
Mahapatra, D.: Neonatal brain mri skull stripping using graph cuts and shape
  priors. In: In Proc: MICCAI workshop on Image Analysis of Human Brain
  Development (IAHBD) (2011)

\bibitem{MahapatraMLMI12}
Mahapatra, D.: Cardiac lv and rv segmentation using mutual context information.
  In: Proc. MICCAI-MLMI. pp. 201--209 (2012)

\bibitem{MahapatraGRSPIE12}
Mahapatra, D.: Groupwise registration of dynamic cardiac perfusion images using
  temporal information and segmentation information. In: In Proc: SPIE Medical
  Imaging (2012)

\bibitem{MahapatraSTACOM12}
Mahapatra, D.: Landmark detection in cardiac mri using learned local image
  statistics. In: Proc. MICCAI-Statistical Atlases and Computational Models of
  the Heart. Imaging and Modelling Challenges (STACOM). pp. 115--124 (2012)

\bibitem{MahapatraJDISkull2012}
Mahapatra, D.: Skull stripping of neonatal brain mri: Using prior shape
  information with graphcuts. J. Digit. Imaging  \textbf{25}(6),  802--814
  (2012)

\bibitem{MahapatraJDIGCSP2013}
Mahapatra, D.: Cardiac image segmentation from cine cardiac mri using graph
  cuts and shape priors. J. Digit. Imaging  \textbf{26}(4),  721--730 (2013)

\bibitem{MahapatraJDIMutCont2013}
Mahapatra, D.: Cardiac mri segmentation using mutual context information from
  left and right ventricle. J. Digit. Imaging  \textbf{26}(5),  898--908 (2013)

\bibitem{MahapatraProISBI13}
Mahapatra, D.: Graph cut based automatic prostate segmentation using learned
  semantic information. In: Proc. IEEE ISBI. pp. 1304--1307 (2013)

\bibitem{MahapatraJDIJSGR2013}
Mahapatra, D.: Joint segmentation and groupwise registration of cardiac
  perfusion images using temporal information. J. Digit. Imaging
  \textbf{26}(2),  173--182 (2013)

\bibitem{MahapatraJDI_Cardiac_FSL}
Mahapatra, D.: An automated approach to cardiac rv segmentation from mri using
  learned semantic information and graph cuts. J. Digit. Imaging.
  \textbf{27}(6),  794--804 (2014)

\bibitem{Mahapatra_LME_CVIU}
Mahapatra, D.: Combining multiple expert annotations using semi-supervised
  learning and graph cuts for medical image segmentation. Computer Vision and
  Image Understanding  \textbf{151}(1),  114--123 (2016)

\bibitem{Mahapatra_OMIA16}
Mahapatra, D.: Retinal image quality classification using neurobiological
  models of the human visual system. In: In Proc. MICCAI-OMIA. pp.~1--8 (2016)

\bibitem{LME_Ar}
Mahapatra, D.: Consensus based medical image segmentation using semi-supervised
  learning and graph cuts. In: arXiv preprint arXiv:1612.02166 (2017)

\bibitem{Mahapatra_LME_PR2017}
Mahapatra, D.: Semi-supervised learning and graph cuts for consensus based
  medical image segmentation. Pattern Recognition  \textbf{63}(1),  700--709
  (2017)

\bibitem{AMD_OCT}
Mahapatra, D.: Amd severity prediction and explainability using image
  registration and deep embedded clustering. In: arXiv preprint
  arXiv:1907.03075 (2019)

\bibitem{GANReg2_Ar}
Mahapatra, D.: Generative adversarial networks and domain adaptation for
  training data independent image registration. In: arXiv preprint
  arXiv:1910.08593 (2019)

\bibitem{DART2020_Ar}
Mahapatra, D.: Registration of histopathogy images using structural information
  from fine grained feature maps. In: arXiv preprint arXiv:2007.02078 (2020)

\bibitem{TMI2021_Ar}
Mahapatra, D.: Interpretability-driven sample selection using self supervised
  learning for disease classification and segmentation. In: arXiv preprint
  arXiv:2104.06087 (2021)

\bibitem{DARTSyn_Ar}
Mahapatra, D.: Learning of inter-label geometric relationships using
  self-supervised learning: Application to gleason grade segmentation. In:
  arXiv preprint arXiv:2110.00404 (2021)

\bibitem{Misc}
Mahapatra, D., Agarwal, K., Khosrowabadi, R., Prasad, D.: Recent advances in
  statistical data and signal analysis: Application to real world diagnostics
  from medical and biological signals. In: Computational and mathematical
  methods in medicine (2016)

\bibitem{MahapatraGAN_ISBI18}
Mahapatra, D., Antony, B., Sedai, S., Garnavi, R.: Deformable medical image
  registration using generative adversarial networks. In: In Proc. IEEE ISBI.
  pp. 1449--1453 (2018)

\bibitem{ISR_Ar}
Mahapatra, D., Bozorgtabar, B.: Retinal vasculature segmentation using local
  saliency maps and generative adversarial networks for image super resolution.
  In: arXiv preprint arXiv:1710.04783 (2017)

\bibitem{PGAN_Ar}
Mahapatra, D., Bozorgtabar, B.: Progressive generative adversarial networks for
  medical image super resolution. In: arXiv preprint arXiv:1902.02144 (2019)

\bibitem{Mahapatra_CMIG2019}
Mahapatra, D., Bozorgtabar, B., Garnavi, R.: Image super-resolution using
  progressive generative adversarial networks for medical image analysis.
  Computerized Medical Imaging and Graphics  \textbf{71},  30--39 (2019)

\bibitem{Mahapatra_CVAMD2021}
Mahapatra, D., Bozorgtabar, B., Ge, Z.: Medical image classification using
  generalized zero shot learning. In: In IEEE CVAMD 2021. pp. 3344--3353 (2021)

\bibitem{Mahapatra_DART21a}
Mahapatra, D., Bozorgtabar, B., Kuanar, S., Ge, Z.: Self-supervised multimodal
  generalized zero shot learning for gleason grading. In: In MICCAI-DART 2021.
  pp. 1--11 (2021)

\bibitem{Mahapatra_CVPR2020}
Mahapatra, D., Bozorgtabar, B., Shao, L.: Pathological retinal region
  segmentation from oct images using geometric relation based augmentation. In:
  In Proc. IEEE CVPR. pp. 9611--9620 (2020)

\bibitem{CVPR2020_Ar}
Mahapatra, D., Bozorgtabar, B., Thiran, J.P., Shao, L.: Pathological retinal
  region segmentation from oct images using geometric relation based
  augmentation. In: arXiv preprint arXiv:2003.14119 (2020)

\bibitem{Mahapatra_MICCAI20}
Mahapatra, D., Bozorgtabar, B., Thiran, J.P., Shao, L.: Structure preserving
  stain normalization of histopathology images using self supervised semantic
  guidance. In: In Proc. MICCAI. pp. 309--319 (2020)

\bibitem{Stain_AR}
Mahapatra, D., Bozorgtabar, B., Thiran, J.P., Shao, L.: Structure preserving
  stain normalization of histopathology images using self supervised semantic
  guidance. In: arXiv preprint arXiv:2008.02101 (2020)

\bibitem{Mahapatra_MICCAI17}
Mahapatra, D., Bozorgtabar, S., Hewavitahranage, S., Garnavi, R.: Image super
  resolution using generative adversarial networks and local saliencymaps for
  retinal image analysis,. In: In Proc. MICCAI. pp. 382--390 (2017)

\bibitem{MahapatraAL_MICCAI18}
Mahapatra, D., Bozorgtabar, S., Thiran, J.P., Reyes, M.: Efficient active
  learning for image classification and segmentation using a sample selection
  and conditional generative adversarial network. In: In Proc. MICCAI (2). pp.
  580--588 (2018)

\bibitem{Mahapatra_OMIA15}
Mahapatra, D., Buhmann, J.: Obtaining consensus annotations for retinal image
  segmentation using random forest and graph cuts. In: In Proc. OMIA. pp.
  41--48 (2015)

\bibitem{Mahapatra_MLMI15_Prostate}
Mahapatra, D., Buhmann, J.: Visual saliency based active learning for prostate
  mri segmentation. In: In Proc. MLMI. pp. 9--16 (2015)

\bibitem{Mahapatra_SSLAL_Pro_JMI}
Mahapatra, D., Buhmann, J.: Visual saliency-based active learning for prostate
  magnetic resonance imaging segmentation. SPIE Journal of Medical Imaging
  \textbf{3}(1),  014003 (2016)

\bibitem{MahapatraRVISBI13}
Mahapatra, D., Buhmann, J.: Automatic cardiac rv segmentation using semantic
  information with graph cuts. In: Proc. IEEE ISBI. pp. 1094--1097 (2013)

\bibitem{MahapatraTIP_RF2014}
Mahapatra, D., Buhmann, J.: Analyzing training information from random forests
  for improved image segmentation. IEEE Trans. Imag. Proc.  \textbf{23}(4),
  1504--1512 (2014)

\bibitem{MahapatraTBME_Pro2014}
Mahapatra, D., Buhmann, J.: Prostate mri segmentation using learned semantic
  knowledge and graph cuts. IEEE Trans. Biomed. Engg.  \textbf{61}(3),
  756--764 (2014)

\bibitem{MahapatraISBI15_Optic}
Mahapatra, D., Buhmann, J.: A field of experts model for optic cup and disc
  segmentation from retinal fundus images. In: In Proc. IEEE ISBI. pp. 218--221
  (2015)

\bibitem{Pat2}
Mahapatra, D., Garnavi, R., Roy, P., Tennakoon, R.: System and method to teach
  and evaluate image grading performance using prior learned expert knowledge
  base. In: US Patent App. 15/459,457 (2018)

\bibitem{Pat3}
Mahapatra, D., Garnavi, R., Roy, P., Tennakoon, R.: System and method to teach
  and evaluate image grading performance using prior learned expert knowledge
  base. In: US Patent App. 15/814,590 (2018)

\bibitem{Pat11}
Mahapatra, D., Garnavi, R., Sedai, S., Roy, P.: Joint segmentation and
  characteristics estimation in medical images. In: US Patent App. 15/234,426
  (2017)

\bibitem{Pat10}
Mahapatra, D., Garnavi, R., Sedai, S., Roy, P.: Retinal image quality
  assessment, error identification and automatic quality correction. In: US
  Patent 9,779,492 (2017)

\bibitem{Pat6}
Mahapatra, D., Garnavi, R., Sedai, S., Tennakoon, R.: Classification of
  severity of pathological condition using hybrid image representation. In: US
  Patent App. 15/426,634 (2018)

\bibitem{Pat4}
Mahapatra, D., Garnavi, R., Sedai, S., Tennakoon, R.: Generating an enriched
  knowledge base from annotated images. In: US Patent App. 15/429,735 (2018)

\bibitem{Pat8}
Mahapatra, D., Garnavi, R., Sedai, S., Tennakoon, R., Chakravorty, R.: Early
  prediction of age related macular degeneration by image reconstruction. In:
  US Patent App. 15/854,984 (2018)

\bibitem{Pat9}
Mahapatra, D., Garnavi, R., Sedai, S., Tennakoon, R., Chakravorty, R.: Early
  prediction of age related macular degeneration by image reconstruction. In:
  US Patent 9,943,225 (2018)

\bibitem{GANReg1_Ar}
Mahapatra, D., Ge, Z.: Combining transfer learning and segmentation information
  with gans for training data independent image registration. In: arXiv
  preprint arXiv:1903.10139 (2019)

\bibitem{Mahapatra_ISBI19}
Mahapatra, D., Ge, Z.: Training data independent image registration with gans
  using transfer learning and segmentation information. In: In Proc. IEEE ISBI.
  pp. 709--713 (2019)

\bibitem{Mahapatra_PR2020}
Mahapatra, D., Ge, Z.: Training data independent image registration using
  generative adversarial networks and domain adaptation. Pattern Recognition
  \textbf{100},  1--14 (2020)

\bibitem{Pat13}
Mahapatra, D., Ge, Z., Sedai, S.: Joint registration and segmentation of images
  using deep learning. In: US Patent App. 16/001,566 (2019)

\bibitem{Mahapatra_MLMI18}
Mahapatra, D., Ge, Z., Sedai, S., Chakravorty., R.: Joint registration and
  segmentation of xray images using generative adversarial networks. In: In
  Proc. MICCAI-MLMI. pp. 73--80 (2018)

\bibitem{Mahapatra_JSTSP2014}
Mahapatra, D., Gilani, S., Saini., M.: Coherency based spatio-temporal saliency
  detection for video object segmentation. IEEE Journal of Selected Topics in
  Signal Processing.  \textbf{8}(3),  454--462 (2014)

\bibitem{MahapatraTMI_CD2013}
Mahapatra, D., J.Tielbeek, Makanyanga, J., Stoker, J., Taylor, S., Vos, F.,
  Buhmann, J.: Automatic detection and segmentation of crohn's disease tissues
  from abdominal mri. IEEE Trans. Med. Imaging  \textbf{32}(12),  1232--1248
  (2013)

\bibitem{MahapatraISBI_CD2014}
Mahapatra, D., J.Tielbeek, Makanyanga, J., Stoker, J., Taylor, S., Vos, F.,
  Buhmann, J.: Active learning based segmentation of crohn's disease using
  principles of visual saliency. In: Proc. IEEE ISBI. pp. 226--229 (2014)

\bibitem{Mahapatra_ABD2014}
Mahapatra, D., J.Tielbeek, Makanyanga, J., Stoker, J., Taylor, S., Vos, F.,
  Buhmann, J.: Combining multiple expert annotations using semi-supervised
  learning and graph cuts for crohn's disease segmentation. In: In Proc:
  MICCAI-ABD (2014)

\bibitem{MahapatraJDICD2013}
Mahapatra, D., J.Tielbeek, Vos, F., Buhmann, J.: A supervised learning approach
  for crohn's disease detection using higher order image statistics and a novel
  shape asymmetry measure. J. Digit. Imaging  \textbf{26}(5),  920--931 (2013)

\bibitem{Mahapatra_DART21b}
Mahapatra, D., Kuanar, S., Bozorgtabar, B., Ge, Z.: Self-supervised learning of
  inter-label geometric relationships for gleason grade segmentation. In: In
  MICCAI-DART 2021. pp. 57--67 (2021)

\bibitem{MahapatraISBI15_JSGR}
Mahapatra, D., Li, Z., Vos, F., Buhmann, J.: Joint segmentation and groupwise
  registration of cardiac dce mri using sparse data representations. In: In
  Proc. IEEE ISBI. pp. 1312--1315 (2015)

\bibitem{MahapatraICIT06}
Mahapatra, D., Routray, A., Mishra, C.: An active snake model for
  classification of extreme emotions. In: IEEE International Conference on
  Industrial Technology (ICIT). pp. 2195--2199 (2006)

\bibitem{Mahapatra_EMBC16}
Mahapatra, D., Roy, P., Sedai, S., Garnavi, R.: A cnn based neurobiology
  inspired approach for retinal image quality assessment. In: In Proc. EMBC.
  pp. 1304--1307 (2016)

\bibitem{Mahapatra_MLMI16}
Mahapatra, D., Roy, P., Sedai, S., Garnavi, R.: Retinal image quality
  classification using saliency maps and cnns. In: In Proc. MICCAI-MLMI. pp.
  172--179 (2016)

\bibitem{MahapatraICBME08_Retrieve}
Mahapatra, D., Roy, S., Sun, Y.: Retrieval of mr kidney images by incorporating
  shape information in histogram of low level features. In: In 13th
  International Conference on Biomedical Engineering. pp. 661--664 (2009)

\bibitem{Pat12}
Mahapatra, D., Saha, S., Vishwanath, A., Bastide, P.: Generating hyperspectral
  image database by machine learning and mapping of color images to
  hyperspectral domain. In: US Patent App. 15/949,528 (2019)

\bibitem{MahapatraTrack_Book}
Mahapatra, D., Saini, M.: A particle filter framework for object tracking using
  visual-saliency information. Intelligent Multimedia Surveillance pp. 133--147
  (2013)

\bibitem{MahapatraICME08}
Mahapatra, D., Saini, M., Sun, Y.: Illumination invariant tracking in office
  environments using neurobiology-saliency based particle filter. In: IEEE
  ICME. pp. 953--956 (2008)

\bibitem{MahapatraMICCAI_CD2013}
Mahapatra, D., Sch$\ddot{u}$ffler, P., Tielbeek, J., Vos, F., Buhmann, J.:
  Semi-supervised and active learning for automatic segmentation of crohn's
  disease. In: Proc. MICCAI, Part 2. pp. 214--221 (2013)

\bibitem{RegGan_Ar}
Mahapatra, D., Sedai, S., Garnavi, R.: Elastic registration of medical images
  with gans. In: arXiv preprint arXiv:1805.02369 (2018)

\bibitem{Pat16}
Mahapatra, D., Sedai, S., Halupka, K.: Uncertainty region based image
  enhancement. In: US Patent App. 10,832,074 (2020)

\bibitem{Covi19_Ar}
Mahapatra, D., Singh, A.: Ct image synthesis using weakly supervised
  segmentation and geometric inter-label relations for covid image analysis.
  In: arXiv preprint arXiv:2106.10230 (2021)

\bibitem{MahapatraMiccai08}
Mahapatra, D., Sun, Y.: Nonrigid registration of dynamic renal {MR} images
  using a saliency based {MRF} model. In: Proc. MICCAI. pp. 771--779 (2008)

\bibitem{MahapatraISBI08}
Mahapatra, D., Sun, Y.: Registration of dynamic renal mr images using
  neurobiological model of saliency. In: Proc. ISBI. pp. 1119--1122 (2008)

\bibitem{MahapatraICBME08_Sal}
Mahapatra, D., Sun, Y.: Using saliency features for graphcut segmentation of
  perfusion kidney images. In: In 13th International Conference on Biomedical
  Engineering (2008)

\bibitem{MahapatraMiccai10}
Mahapatra, D., Sun, Y.: Joint registration and segmentation of dynamic cardiac
  perfusion images using mrfs. In: Proc. MICCAI. pp. 493--501 (2010)

\bibitem{MahapatraICDIP10b}
Mahapatra, D., Sun, Y.: Mrf based joint registration and segmentation of
  dynamic renal mr images. In: Second International Conference on Digital Image
  Processing. vol.~7546, pp. 285--290 (2010)

\bibitem{MahapatraICIP10}
Mahapatra, D., Sun., Y.: An mrf framework for joint registration and
  segmentation of natural and perfusion images. In: Proc. IEEE ICIP. pp.
  1709--1712 (2010)

\bibitem{MahapatraICDIP10a}
Mahapatra, D., Sun, Y.: Retrieval of perfusion images using cosegmentation and
  shape context information. In: Proc. APSIPA Annual Summit and Conference
  (ASC). vol.~35 (2010)

\bibitem{MahapatraEURASIP2010}
Mahapatra, D., Sun, Y.: Rigid registration of renal perfusion images using a
  neurobiology based visual saliency model. EURASIP Journal on Image and Video
  Processing. pp. 1--16 (2010)

\bibitem{MahapatraTBME2011}
Mahapatra, D., Sun, Y.: Mrf based intensity invariant elastic registration of
  cardiac perfusion images using saliency information. IEEE Trans. Biomed.
  Engg.  \textbf{58}(4),  991--1000 (2011)

\bibitem{MahapatraMiccai11}
Mahapatra, D., Sun, Y.: Orientation histograms as shape priors for left
  ventricle segmentation using graph cuts. In: In Proc: MICCAI. pp. 420--427
  (2011)

\bibitem{MahapatraTIP2012}
Mahapatra, D., Sun, Y.: Integrating segmentation information for improved
  mrf-based elastic image registration. IEEE Trans. Imag. Proc.
  \textbf{21}(1),  170--183 (2012)

\bibitem{MahapatraABD12}
Mahapatra, D., Tielbeek, J., Buhmann, J., Vos, F.: A supervised learning based
  approach to detect crohn's disease in abdominal mr volumes. In: Proc. MICCAI
  workshop Computational and Clinical Applications in Abdominal
  Imaging(MICCAI-ABD). pp. 97--106 (2012)

\bibitem{MahapatraCDFssISBI13}
Mahapatra, D., Tielbeek, J., Vos, F., ., J.B.: Crohn's disease tissue
  segmentation from abdominal mri using semantic information and graph cuts.
  In: Proc. IEEE ISBI. pp. 358--361 (2013)

\bibitem{MahapatraCDSPIE13}
Mahapatra, D., Tielbeek, J., Vos, F., Buhmann, J.: Localizing and segmenting
  crohn's disease affected regions in abdominal mri using novel context
  features. In: Proc. SPIE Medical Imaging (2013)

\bibitem{MahapatraWssISBI13}
Mahapatra, D., Tielbeek, J., Vos, F., Buhmann, J.: Weakly supervised semantic
  segmentation of crohn's disease tissues from abdominal mri. In: Proc. IEEE
  ISBI. pp. 832--835 (2013)

\bibitem{MahapatraISBI15_CD}
Mahapatra, D., Vos, F., Buhmann, J.: Crohn's disease segmentation from mri
  using learned image priors. In: In Proc. IEEE ISBI. pp. 625--628 (2015)

\bibitem{Mahapatra_SSLAL_CD_CMPB}
Mahapatra, D., Vos, F., Buhmann, J.: Active learning based segmentation of
  crohns disease from abdominal mri. Computer Methods and Programs in
  Biomedicine  \textbf{128}(1),  75--85 (2016)

\bibitem{MahapatraSPIE08}
Mahapatra, D., Winkler, S., Yen, S.: Motion saliency outweighs other low-level
  features while watching videos. In: SPIE HVEI. pp. 1--10 (2008)

\bibitem{Mahapatra_Thesis}
Mahapatra, D.: Registration and segmentation methodology for perfusion mr
  images: Application to cardiac and renal images. - pp.~-- (2011)

\bibitem{MahapatraTh2012}
Mahapatra, D.: Registration and segmentation methodology for perfusion mr
  images: Application to cardiac and renal images. - pp.~-- (2011)

\bibitem{DARTGZSL_Ar}
Mahapatra, D.: Multimodal generalized zero shot learning for gleason grading
  using self-supervised learning. In: arXiv preprint arXiv:2111.07646 (2021)

\bibitem{IccvGZSl_Ar}
Mahapatra, D.: Generalized zero shot learning for medical image classification.
  In: arXiv preprint arXiv:2204.01728 (2022)

\bibitem{UDA_Ar}
Mahapatra, D.: Unsupervised domain adaptation using feature disentanglement and
  gcns for medical image classification. In: arXiv preprint arXiv:2206.13123
  (2022)

\bibitem{mahapatra2022_midl}
Mahapatra, D., Ge, Z.: {MR} image super resolution by combining feature
  disentanglement {CNN}s and vision transformers. In: Medical Imaging with Deep
  Learning (2022)

\bibitem{ISR_MIDL_Ar}
Mahapatra, D., Ge, Z.: Mr image super resolution by combining feature
  disentanglement cnns and vision transformers. In: - (2022)

\bibitem{MahapatraGZSLTMI}
Mahapatra, D., Ge, Z., Reyes, M.: Self-supervised generalized zero shot
  learning for medical image classification using novel interpretable saliency
  maps. IEEE Transactions on Medical Imaging pp.~1--1 (2022).
  \doi{10.1109/TMI.2022.3163232}

\bibitem{GCN_MIDL_Ar}
Mahapatra, D., Korevaar, S., Tennakoon, R.: Gcn based unsupervised domain
  adaptation with feature disentanglement for medical image classification. In:
  - (2022)

\bibitem{Mahapatra_Media_SIBNET}
Mahapatra, D., Poellinger, A., Reyes, M.: Interpretability-guided inductive
  bias for deep learning based medical image classification and segmentation.
  Medical Image Analysis p. 102551 (2022)

\bibitem{MahapatraTMI2021}
Mahapatra, D., Poellinger, A., Shao, L., Reyes, M.: Interpretability-driven
  sample selection using self supervised learning for disease classification
  and segmentation. IEEE TMI pp. 1--15 (2021)

\bibitem{ZhangCVPR21_27}
Manjón, J.V., Coupé, P., Buades, A., Fonov, V., Collins, D.L.: Non-local mri
  upsampling. Medical Image Analysis  \textbf{14}(6),  1465--1476 (2010)

\bibitem{PandeyiMIMIC2021}
Pandey, A., Paliwal, B., Dhall, A., Subramanian, R., Mahapatra, D.: This
  explains that: Congruent image--report generation for explainable medical
  image analysis with cyclic generative adversarial networks. In: In
  MICCAI-iMIMIC 2021. pp. 1--11 (2021)

\bibitem{SwapVAE}
Park, T., Zhu, J.Y., Wang, O., Lu, J., Shechtman, E., Efros, A.A., Zhang, R.:
  Swapping autoencoder for deep image manipulation. In: Advances in Neural
  Information Processing Systems (2020)

\bibitem{ZhangCVPR21_31}
Pham, C.H., Ducournau, A., Fablet, R., Rousseau, F.: Brain mri super-resolution
  using deep 3d convolutional networks. In: 2017 IEEE 14th International
  Symposium on Biomedical Imaging (ISBI 2017). pp. 197--200 (2017).
  \doi{10.1109/ISBI.2017.7950500}

\bibitem{ZhangCVPR21_32}
Plenge, E., et~al: Super-resolution methods in mri: can they improve the
  trade-off between resolution, signal-to-noise ratio, and acquisition time?
  Magnetic resonance in medicine  \textbf{68}(6),  1983--1993 (2012)

\bibitem{Roy_DICTA16}
Roy, P., Chakravorty, R., Sedai, S., Mahapatra, D., Garnavi, R.: Automatic eye
  type detection in retinal fundus image using fusion of transfer learning and
  anatomical features. In: In Proc. DICTA. pp.~1--7 (2016)

\bibitem{Roy_ISBI17}
Roy, P., Tennakoon, R., Cao, K., Sedai, S., Mahapatra, D., Maetschke, S.,
  Garnavi, R.: A novel hybrid approach for severity assessment of diabetic
  retinopathy in colour fundus images,. In: In Proc. IEEE ISBI. pp. 1078--1082
  (2017)

\bibitem{Pat18}
Roy, P., Mahapatra, D., Garnavi, R., Tennakoon, R.: System and method to teach
  and evaluate image grading performance using prior learned expert knowledge
  base. In: US Patent App. 10,984,674 (2021)

\bibitem{sZoom_Ar}
Saini, M., Guthier, B., Kuang, H., Mahapatra, D., Saddik, A.: szoom: A
  framework for automatic zoom into high resolution surveillance videos. In:
  arXiv preprint arXiv:1909.10164 (2019)

\bibitem{Schuffler_ABD2013}
Sch$\ddot{u}$ffler, P., Mahapatra, D., Tielbeek, J., Vos, F., Makanyanga, J.,
  Pends, D., Nio, C., Stoker, J., Taylor, S., Buhmann, J.: A model development
  pipeline for crohns disease severity assessment from magnetic resonance
  images. In: In Proc: MICCAI-ABD (2013)

\bibitem{Schuffler_ABD2014}
Sch$\ddot{u}$ffler, P., Mahapatra, D., Tielbeek, J., Vos, F., Makanyanga, J.,
  Pends, D., Nio, C., Stoker, J., Taylor, S., Buhmann, J.: Semi automatic
  crohns disease severity assessment on mr imaging. In: In Proc: MICCAI-ABD
  (2014)

\bibitem{ZhangCVPR21_36}
Scherrer, B., Gholipour, A., Warfield, S.K.: Super-resolution reconstruction to
  increase the spatial resolution of diffusion weighted images from orthogonal
  anisotropic acquisitions. Medical Image Analysis  \textbf{16}(7),  1465--1476
  (2012)

\bibitem{Schuffler_ABD2014_2}
Schüffler, P.J., Mahapatra, D., Vos, F.M., Buhmann, J.M.: Computer aided
  crohn’s disease severity assessment in mri. In: VIGOR++ Workshop
  2014-Showcase of Research Outcomes and Future Outlook. pp.~-- (2014)

\bibitem{Sedai_OMIA18}
Sedai, S., Mahapatra, D., Antony, B., Garnavi, R.: Joint segmentation and
  uncertainty visualization of retinal layers in optical coherence tomography
  images using bayesian deep learning. In: In Proc. MICCAI-OMIA. pp. 219--227
  (2018)

\bibitem{Sedai_MLMI18}
Sedai, S., Mahapatra, D., Ge, Z., Chakravorty, R., Garnavi, R.: Deep multiscale
  convolutional feature learning for weakly supervised localization of chest
  pathologies in x-ray images. In: In Proc. MICCAI-MLMI. pp. 267--275 (2018)

\bibitem{Sedai_MICCAI17}
Sedai, S., Mahapatra, D., Hewavitharanage, S., Maetschke, S., Garnavi, R.:
  Semi-supervised segmentation of optic cup in retinal fundus images using
  variational autoencoder,. In: In Proc. MICCAI. pp. 75--82 (2017)

\bibitem{Sedai_EMBC16}
Sedai, S., Roy, P., Mahapatra, D., Garnavi, R.: Segmentation of optic disc and
  optic cup in retinal fundus images using shape regression. In: In Proc. EMBC.
  pp. 3260--3264 (2016)

\bibitem{Sedai_OMIA16}
Sedai, S., Roy, P., Mahapatra, D., Garnavi, R.: Segmentation of optic disc and
  optic cup in retinal images using coupled shape regression. In: In Proc.
  MICCAI-OMIA. pp.~1--8 (2016)

\bibitem{SrivastavaFAIR2021}
Srivastava, S., Yaqub, M., Nandakumar, K., Ge, Z., Mahapatra, D.: Continual
  domain incremental learning for chest x-ray classification in low-resource
  clinical settings. In: In MICCAI-FAIR 2021. pp. 1--11 (2021)

\bibitem{Tennakoon_OMIA16}
Tennakoon, R., Mahapatra, D., Roy, P., Sedai, S., Garnavi, R.: Image quality
  classification for dr screening using convolutional neural networks. In: In
  Proc. MICCAI-OMIA. pp. 113--120 (2016)

\bibitem{TongDART20}
Tong, J., Mahapatra, D., Bonnington, P., Drummond, T., Ge, Z.: Registration of
  histopathology images using self supervised fine grained feature maps. In: In
  Proc. MICCAI-DART Workshop. pp. 41--51 (2020)

\bibitem{Unetr41}
Valanarasu, J.M.J., Oza, P., Hacihaliloglu, I., Patel, V.M.: Medical
  transformer: Gated axial-attention for medical image segmentation (2021)

\bibitem{MonusacTMI}
Verma, R., Kumar, N., Patil, A., Kurian, N.C., Rane, S., Graham, S., Vu, Q.D.,
  Zwager, M., Raza, S.E.A., Rajpoot, N., Wu, X., Chen, H., Huang, Y., Wang, L.,
  Jung, H., Brown, G.T., Liu, Y., Liu, S., Jahromi, S.A.F., Khani, A.A.,
  Montahaei, E., Baghshah, M.S., Behroozi, H., Semkin, P., Rassadin, A.,
  Dutande, P., Lodaya, R., Baid, U., Baheti, B., Talbar, S., Mahbod, A., Ecker,
  R., Ellinger, I., Luo, Z., Dong, B., Xu, Z., Yao, Y., Lv, S., Feng, M., Xu,
  K., Zunair, H., Hamza, A.B., Smiley, S., Yin, T.K., Fang, Q.R., Srivastava,
  S., Mahapatra, D., Trnavska, L., Zhang, H., Narayanan, P.L., Law, J., Yuan,
  Y., Tejomay, A., Mitkari, A., Koka, D., Ramachandra, V., Kini, L., Sethi, A.:
  Monusac2020: A multi-organ nuclei segmentation and classification challenge.
  IEEE Transactions on Medical Imaging  \textbf{40}(12),  3413--3423 (2021).
  \doi{10.1109/TMI.2021.3085712}

\bibitem{VosEMBC}
Vos, F.M., Tielbeek, J., Naziroglu, R., Li, Z., Sch$\ddot{u}$ffler, P.,
  Mahapatra, D., Wiebel, A., Lavini, C., Buhmann, J., Hege, H., Stoker, J., van
  Vliet, L.: Computational modeling for assessment of {IBD}: to be or not to
  be? In: Proc. IEEE EMBC. pp. 3974--3977 (2012)

\bibitem{UNETR_44}
Wang, W., Xie, E., Li, X., Fan, D.P., Song, K., Liang, D., Lu, T., Luo, P.,
  Shao, L.: Pyramid vision transformer: A versatile backbone for dense
  prediction without convolutions. In: Proceedings of the IEEE/CVF
  International Conference on Computer Vision (ICCV). pp. 568--578 (October
  2021)

\bibitem{Unetr43}
Wang, W., Chen, C., Ding, M., Li, J., Yu, H., Zha, S.: Transbts: Multimodal
  brain tumor segmentation using transformer (2021)

\bibitem{Unetr47}
Xie, Y., Zhang, J., Shen, C., Xia, Y.: Cotr: Efficiently bridging cnn and
  transformer for 3d medical image segmentation (2021)

\bibitem{Xing_MICCAI19}
Xing, Y., Ge, Z., Zeng, R., Mahapatra, D., Seah, J., Law, M., Drummond, T.:
  Adversarial pulmonary pathology translation for pairwise chest x-ray data
  augmentation. In: In Proc. MICCAI. pp. 757--765 (2019)

\bibitem{ZhangCVPR21_44}
Xu, K., Ba, J., Kiros, R., Cho, K., Courville, A., Salakhutdinov, R., Zemel,
  R., Bengio, Y.: Show, attend and tell: Neural image caption generation with
  visual attention (2016)

\bibitem{FengMulti_21}
Xuan, K., Sun, S., Xue, Z., Wang, Q., Liao, S.: Learning mri k-space
  subsampling pattern using progressive weight pruning. In: Martel, A.L.,
  Abolmaesumi, P., Stoyanov, D., Mateus, D., Zuluaga, M.A., Zhou, S.K.,
  Racoceanu, D., Joskowicz, L. (eds.) Medical Image Computing and Computer
  Assisted Intervention -- MICCAI 2020. pp. 178--187. Springer International
  Publishing, Cham (2020)

\bibitem{Fastmri}
Zbontar, J., Knoll, F., Sriram, A., Murrell, T., Huang, Z., Muckley, M.J.,
  Defazio, A., Stern, R., Johnson, P., Bruno, M., Parente, M., Geras, K.J.,
  Katsnelson, J., Chandarana, H., Zhang, Z., Drozdzal, M., Romero, A., Rabbat,
  M., Vincent, P., Yakubova, N., Pinkerton, J., Wang, D., Owens, E., Zitnick,
  C.L., Recht, M.P., Sodickson, D.K., Lui, Y.W.: fastmri: An open dataset and
  benchmarks for accelerated mri (2019)

\bibitem{ZhangCVPR21}
Zhang, Y., Li, K., Li, K., Fu, Y.: Mr image super-resolution with squeeze and
  excitation reasoning attention network. In: 2021 IEEE/CVF Conference on
  Computer Vision and Pattern Recognition (CVPR). pp. 13420--13429 (2021).
  \doi{10.1109/CVPR46437.2021.01322}

\bibitem{ZhangCVPR21_48}
Zhang, Y., Tian, Y., Kong, Y., Zhong, B., Fu, Y.: Residual dense network for
  image super-resolution. In: CVPR (2018)

\bibitem{FengMulti_25}
Zhao, C., Carass, A., Dewey, B., Woo, J., Oh, J., Calabresi, P., Reich, D.,
  Sati, P., Pham, D., Prince, J.: A deep learning based anti-aliasing self
  super-resolution algorithm for mri. In: Schnabel, J., Davatzikos, C.,
  Alberola-L{\'o}pez, C., Fichtinger, G., Frangi, A. (eds.) Medical Image
  Computing and Computer Assisted Intervention – MICCAI 2018 - 21st
  International Conference, 2018, Proceedings. pp. 100--108. Springer Verlag
  (2018)

\bibitem{ZhangCVPR21_49}
Zhao, X., Zhang, Y., Zhang, T., Zou, X.: Channel splitting network for single
  mr image super-resolution. IEEE Transactions on Image Processing
  \textbf{28}(11),  5649--5662 (2019)

\bibitem{Unetr52}
Zheng, S., Lu, J., Zhao, H., Zhu, X., Luo, Z., Wang, Y., Fu, Y., Feng, J.,
  Xiang, T., Torr, P.H.S., Zhang, L.: Rethinking semantic segmentation from a
  sequence-to-sequence perspective with transformers (2021)

\bibitem{ZhouTMI20}
Zhou, T., Fu, H., Chen, G., Shen, J., Shao, L.: Hi-net: Hybrid-fusion network
  for multi-modal mr image synthesis. IEEE Transactions on Medical Imaging
  \textbf{39}(9),  2772--2781 (2020). \doi{10.1109/TMI.2020.2975344}

\bibitem{UNETR_55}
Zhu, X., Su, W., Lu, L., Li, B., Wang, X., Dai, J.: Deformable detr: Deformable
  transformers for end-to-end object detection (2021)

\bibitem{Mahapatra_MLMI15_Optic}
Zilly, J., Buhmann, J., Mahapatra, D.: Boosting convolutional filters with
  entropy sampling for optic cup and disc image segmentation from fundus
  images. In: In Proc. MLMI. pp. 136--143 (2015)

\bibitem{Zilly_CMIG_2016}
Zilly, J., Buhmann, J., Mahapatra, D.: Glaucoma detection using entropy
  sampling and ensemble learning for automatic optic cup and disc segmentation.
  In Press Computerized Medical Imaging and Graphics  \textbf{55}(1),  28--41
  (2017)

\end{thebibliography}

\end{document}